\documentclass{myaastex}
\usepackage{emulateapj5}
\newcommand{\mr}[1]{\mathrm{#1}}

\newcommand{\mc}[1]{\multicolumn{2}{c}{{#1}}}

\newcommand{\m}{$^{-1}$}

\bibpunct{(}{)}{;}{a}{}{,}%

\submitted{Submited September 30, 2004, accepted for publication in
the Astrophysical Journal on December 8, 2004}

\begin{document}

\title{XMM-Newton and Gemini Observations of Eight RASSCALS 
Galaxy Groups\altaffilmark{1}}

\altaffiltext{1}{ Based on observations obtained with XMM-Newton, an
ESA science mission with instruments and contributions directly funded
by ESA Member States and the USA (NASA). The XMM-Newton project is
supported by the Bundesministerium f\"{u}r Bildung und
Forschung/Deutsches Zentrum f\"{u}r Luft- und Raumfahrt (BMFT/DLR),
the Max-Planck Society and the Heidenhain-Stiftung, and also by PPARC,
CEA, CNES, and ASI. Also based on observations obtained at the Gemini
Observatory, which is operated by the Association of Universities for
Research in Astronomy, Inc., under a cooperative agreement with the
NSF on behalf of the Gemini partnership: the National Science
Foundation (United States), the Particle Physics and Astronomy
Research Council (United Kingdom), the National Research Council
(Canada), CONICYT (Chile), the Australian Research Council
(Australia), CNPq (Brazil), and CONICET (Argentina). }

\author{Andisheh Mahdavi}
\affil{Institute for Astronomy, University of Hawaii}

\author{Alexis Finoguenov and Hans B\"ohringer}
\affil{Max-Planck-Institut f\"ur Extraterrestrische Physik}

\author{Margaret J. Geller}
\affil{Harvard-Smithsonian Center for Astrophysics}

\and

\author{J. Patrick Henry}
\affil{Institute for Astronomy, University of Hawaii}

\newcommand{\fitalphap}{$-0.78^{+0.04}_{-0.03}$}
\newcommand{\fitgammap}{$-1.7^{+0.1}_{-0.3}$}
\newcommand{\fitalphas}{$0.92^{+0.04}_{-0.05}$}
\newcommand{\fitgammas}{$0.42^{+0.05}_{-0.04}$}

\begin{abstract}
We study the distribution of gas pressure and entropy in eight groups
of galaxies belonging to the ROSAT All-Sky Survey / Center for
Astrophysics Loose Systems (RASSCALS). We use archival and proprietary
XMM-Newton observations, supplementing the X-ray data with redshifts
derived from the literature; we also list 127 new redshifts measured
with the Gemini North telescope. The groups are morphologically
heterogeneous in both the optical and the X-ray, and several suffer
from superpositions with background galaxies or clusters of galaxies.
Nevertheless, they show remarkable self-similarity in their
azimuthally averaged entropy and temperature profiles. The entropy
increases with radius; the behavior of the entropy profiles is
consistent with an increasing broken power law with inner and outer
slope \fitalphas\ and \fitgammas\ (68\% confidence),
respectively. There is no evidence of a central, isentropic core, and
the entropy distribution in most of the groups is flatter at large
radii than in the inner region, challenging earlier reports as well as
theoretical models predicting large isentropic cores or asymptotic
slopes of 1.1 as $r \rightarrow \infty$. The pressure profiles are
consistent with a self-similar decreasing broken power law in radius;
the inner and outer slopes are \fitalphap\ and \fitgammap,
respectively. The results suggest that the larger scatter in the
entropy distribution reflects the varied gasdynamical histories of the
groups; the regularity and self-similarity of the pressure profiles is
a sign of a similarity in the underlying dark matter distributions.
\end{abstract}

\section{Introduction}

Groups are the intermediate structures between galaxies and clusters
of galaxies, and thus are crucial to an understanding of how galaxies
evolve in dense environments
\nocite{Zwicky37,Zwicky60,Hickson82,Geller83,Ramella89,White99}({Zwicky} 1937; {Zwicky} \& {Humason} 1960; {Hickson} 1982; {Geller} \& {Huchra} 1983; {Ramella}, {Geller}, \& {Huchra} 1989; {White} {et~al.} 1999). In the
X-ray, groups and clusters are customarily distinguished by the
temperature of the hot, gaseous medium that makes up $\approx 10\%$
of their total mass
\nocite{Ebeling94,Ponman96,Mahdavi00,Mulchaey00}({Ebeling}, {Voges}, \&  {B\"ohringer} 1994; {Ponman} {et~al.} 1996; {Mahdavi} {et~al.} 2000; {Mulchaey} 2000). Clusters, with
temperatures $\gtrsim 2$ keV, contribute only about 3\% to the matter
density of the Universe; but groups, with X-ray temperatures in the
range 0.5-2.0 keV (corresponding to a mass range $10^{13}$--$10^{14}
M_\odot$), contribute about twice as much \nocite{Ikebe02,Reiprich02}({Ikebe} {et~al.} 2002; {Reiprich} \& {B{\" o}hringer} 2002).
Because of the cooler intracluster medium (ICM) temperature of the
groups, the set of observable X-ray spectral lines is richer than it
is in clusters, and allows a more accurate determination of elemental
abundances.  
The cD galaxies in groups make up a larger fraction of the total
baryonic mass of the system than do the cD galaxies in clusters
\nocite{Helsdon01,Lin04}({Helsdon} {et~al.} 2001; {Lin} \& {Mohr} 2004); hence the properties of the gaseous medium
are more directly affected by the formation and evolution of the
central galaxy. Finally, because the typical velocities of galaxies in
groups are comparable to the galaxies' internal (stellar) velocity
dispersions, galaxy interactions through the merging instability are
common and very effective \nocite{Diaferio93,Ramella94}({Diaferio} {et~al.} 1993; {Ramella} {et~al.} 1994).

Recent studies suggest that in the X-ray, the distinction between
groups and clusters goes beyond an arbitrary temperature or mass
boundary. Differences in the physical state of the intragroup medium
make it difficult to view the $\lesssim 2$ keV systems simply as
smaller, ``rescaled'' versions of the more massive clusters. Of
particular interest are the heating of the ICM by nongravitational
processes such as supernova explosions, stellar winds, AGN activity,
and shocks resulting from interaction with the surrounding large scale
structure \nocite{Ponman99,Loewenstein01,Tornatore03}({Ponman}, {Cannon}, \& {Navarro} 1999; {Loewenstein} 2001; {Tornatore} {et~al.} 2003). Sometimes called
``preheating,'' these processes can leave a distinct mark on the
entropy distribution of the gas in groups, to a degree not observable
in more massive clusters. For this reason groups provide a fossil
record of this energy production during cosmic structure formation and
galaxy evolution.

In this study we focus on the distribution of entropy and pressure in
eight systems drawn from the RASSCALS survey of nearby galaxy
groups. In \S\ref{sec:data} and \S\ref{sec:xmm} we discuss the optical
and X-ray data we have gathered for this study, as well as the data
reduction procedures used in our analysis. In \S\ref{sec:profiles} we
describe our attempts to fit self-similar profiles to the entropy and
the pressure distribution.  In \S\ref{sec:individual} we describe each
system in detail, In \S\ref{sec:conclusion} we summarize our
conclusions.

\section{Optical Data}
\label{sec:data}

\subsection{Sample Selection}

At the focus of our study are groups with emission-weighted
temperature $\sim 0.5-2$ keV. The X-ray luminosity-temperature
relation \nocite{Ikebe02}({Ikebe} {et~al.} 2002) suggests that these groups should have $L_X
\lesssim 10^{43}$ erg s\m\ in the 0.1-2.4 keV energy range.\footnote{We
assume $H_0 = 70$ km s\m\ Mpc\m, $\Omega_0 = 0.3$, and
$\Omega_\Lambda=0.7$ throughout the paper.}

To build our sample, we begin with the ROSAT All-Sky Survey-Center for
Astrophysics Loose Systems (RASSCALS), a statistically complete,
magnitude-limited catalog of optically identified groups with $0.01 <
z < 0.04$ \nocite{Mahdavi00}({Mahdavi} {et~al.} 2000). This catalog contains 260 groups, of
which 43 have X-ray emission detected in the RASS. To conduct a more
detailed study of the properties of these groups, we select X-ray
emitting RASSCALS with $L_X < 10^{43}$ erg s\m\ that have been
observed with the XMM-Newton observatory. The properties of the
sample are shown in Table~\ref{tbl:sample}.

The membership of the original RASSCALS groups was established
spectroscopically with a completeness limit $m_R \approx 14.4$
\nocite{Ramella02}({Ramella} {et~al.} 2002). All but one of the groups we have selected for
analysis have also been the targets of deeper redshift surveys, by
\nocite{Mahdavi04}{Mahdavi} \& {Geller} (2004) and \nocite{Rines03}{Rines} {et~al.} (2003) (complete to $m_R \approx 15.4$),
by \nocite{Zabludoff98}{Zabludoff} \& {Mulchaey} (1998) (70\%--95\% complete to different magnitude
limits depending on the group), or by \nocite{Pinkney93}{Pinkney} {et~al.} (1993) (unknown
completeness). None of the groups had significant overlap with the
current release of the Sloan Digital Sky Survey \nocite{Abazajian04}({Abazajian} {et~al.} 2004).
In addition, we have performed optical observations which we describe
below.

\subsection{New Optical Spectroscopy}

We observed three groups, RGH 80, HCG 97, and NRGb184 with the 8m
Gemini North telescope on Mauna Kea, Hawaii. The Gemini multi-object
spectrograph, GMOS, was used to measure redshifts for galaxies as
faint as $m_R \approx 20$ within 6.5\arcmin\ of the center of each
system.

The first step was the selection of targets for spectroscopy. We used
GMOS in queue imaging mode to obtain a $2\times2$, $13\arcmin \times
13\arcmin$ $R$-band mosaic around the central galaxy in each group;
the length of each imaging exposure was 10 minutes. The SExtractor
package \nocite{Bertin96}({Bertin} \& {Arnouts} 1996) then separated galaxies from stars, and
calculated the magnitudes and half-light radii of the galaxies. It
would have taken an inordinate amount of time to measure redshifts for
all $m_R < 20$ galaxies in each image. Therefore, galaxies were sorted
by their half-light radii, with the assumption that regardless of
magnitude, galaxies with larger half-light radii would be more likely
to have low redshifts and therefore belong to the group. All three
groups are at a small enough redshift that foreground contamination
was not an issue with this selection procedure.

In each piece of the mosaic, we selected the 20 galaxies with the
largest half-light radii and without previously measured
redshifts. Thus there were a total of 80 targets for NRGb184 and HCG
97. For RGH80, we included an additional central mask, for a total of
100 target galaxies. We designed one slit mask per image to be used
with the GMOS B600 grating; the slit size was 0.5\arcsec (RGH80 and
HCG 97) or 0.75\arcsec (NRGb184) achieving a resolution of $4-6\AA$ at
$4000\AA$. The wavelength coverage was $4000-6000\AA$, with a shift of
$0-1000\AA$ in either direction depending on the position of the
galaxy on the focal plane.  We used two 30 minute exposures per slit
mask (a total of one hour), taking arc lamp exposures between each set
of two exposures to obtain accurate wavelength calibrations. Because
absolute photometric calibration is not required for redshift
measurements, we did not apply flat-fielding or flux calibrations to
the spectra.

The Gemini GMOS package for IRAF\footnote{For a description of the
Gemini data reduction package see
\url{http://www.gemini.edu/sciops/data/dataSoftware.html}.}  was used
to calculate the wavelength solutions and to reduce the multi-object
observations into one-dimensional spectra. The RVSAO package
\nocite{Kurtz98}({Kurtz} \& {Mink} 1998), incorporating the methods of \nocite{TonryDavis79}{Tonry} \& {Davis} (1979),
allowed us to measure redshifts by maximizing the cross-correlation of
the spectra with absorption- and emission-line templates. The
estimation of errors in the optical velocities is described in detail
by \nocite{Kurtz98}{Kurtz} \& {Mink} (1998), who determine the relationship between the shape of
the cross-correlation function peak and the 68\% velocity confidence
interval using galaxies with known velocities.  Our results are in
Table~\ref{tbl:gemini}.

\subsection{Membership}

We assemble a galaxy catalog by combining all the available redshift
surveys in the direction of each group, and restricting ourselves to
galaxies within 2 Mpc of the group center. The
mean redshift $z$ and velocity dispersion $\sigma$ of each group are
defined as follows \nocite{Danese80}({Danese}, {de Zotti}, \& {di  Tullio} 1980):
\begin{eqnarray}
z & \equiv & \frac{1}{N} \sum_{i=1}^N z_i; \\
\sigma^2 & \equiv & \sum_{i=1}^N \frac{(c z_i - c z)^2 - 
\epsilon_i^2}{(N-1)(1+z)^2}.
\end{eqnarray}
The subtraction of the root-mean-square velocity error in the second
term is an attempt to remove the contribution of the measurement
uncertainties $\epsilon_i$ to $\sigma$. In practice, this correction
may be inaccurate when the underlying velocity distribution is
nongaussian, the sample size is small ($N \lesssim 20$), and the
velocity dispersion is of the same order of magnitude as
$\epsilon_i$. The inaccuracy is not a source of concern for our
sample, but may be relevant for some extremely low-velocity dispersion
groups with $\sigma \approx 70$ km s\m\ \nocite{Mahdavi99,Mahdavi00}({Mahdavi} {et~al.} 1999, 2000).

To determine group membership, we then use the ``sigma-clipping''
\nocite{Zabludoff90, Mahdavi04}({Zabludoff}, {Huchra}, \&  {Geller} 1990; {Mahdavi} \& {Geller} 2004) algorithm to reject outliers, or
galaxies unlikely to be bound to the group. This algorithm consists
entirely of making sure that no group member is separated from its
nearest neighbor in velocity space by more than the velocity
dispersion $\sigma$ of the group.

\section{XMM-Newton Observations}
\label{sec:xmm}

We conducted XMM-Newton Observations of three of the eight groups,
while the remaining five are publicly available in the XMM-Newton Data
Archive. Table~\ref{tbl:xmmprops} contains the details of the
observations.

The initial steps of the data reduction are similar to the procedure
described in \nocite{Zhang04}{Zhang} {et~al.} (2004) and \nocite{Finoguenov03}{Finoguenov} {et~al.} (2003).  The first
important aspect is the removal of flares, which can significantly
enhance the detector background, severely limiting the detection of
low surface brightness features. Thus, for the group analysis, using
flare free observing periods is critical.  At energies above 10 keV
the particle background dominates the detected counts; there is little
X-ray emission from our targets because (1) the telescope efficiency
is quite low at energies $>10$ keV, and (2) the temperature of the
objects in our study is less than 2 keV.  We use the 10--15 keV energy
band (binned in 100s intervals) to monitor the particle background
and to excise periods of high particle flux. In this screening process
we use the settings FLAG=0 and PATTERN$<5$ for the pn detector on
XMM-Newton. We reject time intervals affected by flares by
excising periods where the detector count rate exceeds the mean
quiescent rate by more than 2$\sigma$. To produce the broad-band
images and hardness ratio maps, we have also used the MOS1 and MOS2
events with PATTERN$<13$ and FLAG=0.  To reduce the widths of the gaps
in the pn broad band images, we included photons near the pn-CCD
borders, near bad pixels, and near offset columns. Data from the pn
detector were used exclusively in the spectroscopic analysis described
below.

Because most of the observations analyzed here were performed using a
short integration frame time for pn (Full Frame Mode), it is important
to remove the out-of-time events (OOTE) for accurate imaging and
spectral analysis. We used the standard product of the XMMSAS 5.4 {\it
epchain} task to produce the simulated OOTE file for all the
observations and scale it by the fraction of the OOTE expected for the
frame exposure time, (Table~\ref{tbl:xmmprops}). 

The cosmic component of the background consists of emission from the
Galaxy as well as the extragalactic Cosmic X-ray Background
(CXB). Observations of blank fields also contain both
components. Provided that the exposures are done with the same
instrumental set-up (e.g. with a particular filter) the spectra of the
CXB are the same for both the target and the blank field.  The
Galactic component acts as an absorber and emitter, thus it is
important to choose a similar absorbing column density for both the
target and background data sets, so that the expected background
spectra are similar. In addition, there are variations in the Galactic
emission on small scales.

The vignetting correction is performed taking into account the source
extent, recent vignetting calibration \nocite{Lumb04}({Lumb} {et~al.} 2004), and the pn
response matrices released under XMMSAS 6.0. The residual systematic
uncertainty of the flux is below 4\% for the pn \nocite{Lumb04}({Lumb} {et~al.} 2004).
\footnote{ For further details of XMM-Newton processing we refer the
reader to \url{http://wave.xray.mpe.mpg.de/xmm/cookbook/general}.}

The analysis of each group consists of two steps: (1) construction of
separate wavelet-decomposed maps of estimated surface brightness in
the 0.5-2.0 keV band ($I$) and of emission-weighted temperature
($T_I$), which may be used to derive the entropy and pressure
``integrated'' along the line of sight; and (2) use of the
wavelet-decomposed maps to identify contiguous regions from which we
extract X-ray spectra for independent model-fitting.

The background-subtracted wavelet maps are based on photon images
corrected for instrumental effects. The surface brightness map is
constructed using the technique described in \nocite{Vikhlinin98}{Vikhlinin} {et~al.} (1998). A
hardness map of the emission then results from dividing the
wavelet-reconstructed images in the 0.5--1 and 1--2 keV bands. The
hardness ratio is a monotonic and unique function of the
emission-weighted temperature, as long as the group redshift is well
known (true for our sample). The advantages of using wavelets include
the ability to remove additional background by spatial filtering and a
control over the statistical significance of the detected
structures. We use the ``a trous'' method of wavelet image
reconstruction with scales from $8^{\prime\prime}$ to $4^\prime$,
applying a four sigma detection threshold and retaining emission to a
$1.7\sigma$ detection limit. Occasionally, the wavelet algorithm
generates small scale discontinuities in the reconstructed image. We
remedy this effect by applying additional smoothing before producing
the hardness ratio maps.

We also construct maps of the entropy and pressure ``integrated''
along the line of sight, using the surface brightness image and the
emission-weighted temperature maps. The use of the word ``integrated''
is not strict, because these quantities do not have the dimensions of
entropy or pressure times distance, and they serve as heuristic aids
only. In studies of the intracluster medium, the common definitions of
entropy and pressure are $k T / n^{2/3}$ and $ n k T$, respectively,
where $n$ is the gas density and $k T$ is the temperature in units of
keV (see \S\ref{sec:profiles} for further details). Because the
emission measure is proportional to $n^2$, we use $\sqrt{I}$ as an
estimate of the gas density ``integrated'' along the line of
sight. The ``integrated'' entropy is calculated simply as $T_I
I^{-1/3}$, and the corresponding pressure as $\sqrt{I} T_I$. The
quantities shown in these maps are not simply related to the physical,
unprojected entropy and pressure, and they suffer from a number of
degeneracies, with metallicity-density being the strongest (a
significant fraction of the group emission results from line
emission). Nevertheless, the maps indicate the regions of primary
interest for further spectroscopic analysis, in which most of the
degeneracies are removed. In general, we expect the entropy to
increase monotonically with radius \nocite{Metzler94}({Metzler} \& {Evrard} 1994), while the
pressure should decrease. We stress that the ``integrated'' pressure
and entropy are used only as tools to guide further spectroscopic
analysis. They are not used in the fits we discuss in
\S\ref{sec:profiles}.

The second, spectroscopic part of the analysis uses a mask file,
constructed from both hardness ratio and surface brightness analysis
described above. The first application of this technique is in
\nocite{Finoguenov04}{Finoguenov} {et~al.} (2004). In our analysis we select regions with similar
spectral properties.  We combine the regions so that counting
statistics are not the limiting factor in our derivation of the group
properties.  We use the wavelet-based maps to identify regions with
similar X-ray colors and intensity levels. To generate the mask file
for use in the further spectral analysis, we sample the changes in the
intensity and hardness ratio at the precision allowed by the
statistics of the data. We then examine each of the isolated regions
with approximately equal color and intensity, imposing the additional
criterion that the reg ions should be larger than the PSF width
(15\arcsec ) and contain more than 300 counts in the raw pn image. A
sample mask file is shown in Figure \ref{fig:region}.

The spectral analysis was performed using single-temperature models
and the APEC plasma code, fitting elemental abundances from O to Ni as
one group and assuming the photospheric solar abundance ratios of
\nocite{Anders89}({Anders} \& {Grevesse} 1989). It should be noted that evidence for multiphase gas
in groups of galaxies exists \nocite{Buote03}({Buote} {et~al.} 2003), such that gas at each
\emph{projected} radius may vary in temperature by a factor of
$\approx 2$; however, if the variations are due to a smooth radial
trend in the plasma temperature (i.e., a temperature profile), a
series of single-temperature fits across the group should roughly
reproduce the general shape of the underlying pressure and entropy
distributions.

In our analysis, we paid special attention to the issue of background
estimation. We employ a double background subtraction technique,
following \nocite{Zhang04}{Zhang} {et~al.} (2004).  We used the region with radius
12--16$^\prime$ and 2--12 keV band to estimate the quality of the
background subtraction using blank fields. This region is free of
significant X-ray emission from the groups.  We fix the shape of the
residual background component, and also add a 0.2 keV thermal
component (APEC with solar element abundance) to account for a
possible variation in the Galactic foreground, allowing the
normalizations of both components to be fit.

To estimate the true (as opposed to ``integrated'') pressure $n k T$
and entropy $ k T n^{-2/3}$ in each region, we need measurements of
both the gas density $n$ and the temperature $T$. The spectral fit
products, however, are $k T$ and $n^2 V$, where $V$ is the volume of
the emitting region. We do not have an exact knowledge of $V$. Because
the mask generation technique identifies interesting regions by their
wavelet-decomposed properties---producing region maps like Figure
\ref{fig:region}---a straightforward deprojection of the spectra is
not possible.  We therefore need to estimate the length of the column
for each selected two-dimensional region on the sky. We assume that
the gas emits uniformly along the line of sight as shown in Figure
\ref{fig:geometry}. With these approximations, the longest length
through each volume is $L=2 \sqrt{R_2^2 - R_1^2}$, and the mean three
dimensional distance of the region from the cluster center is $r =
(R_1+R_2)/2$. The volume of the region is then $2 S L / 3$, where S is
the area of the region. A similar procedure is used in \nocite{Henry04}{Henry}, {Finoguenov}, \& {Briel} (2004).

\begin{tabular}{c}
\resizebox{3.5in}{!}{\includegraphics{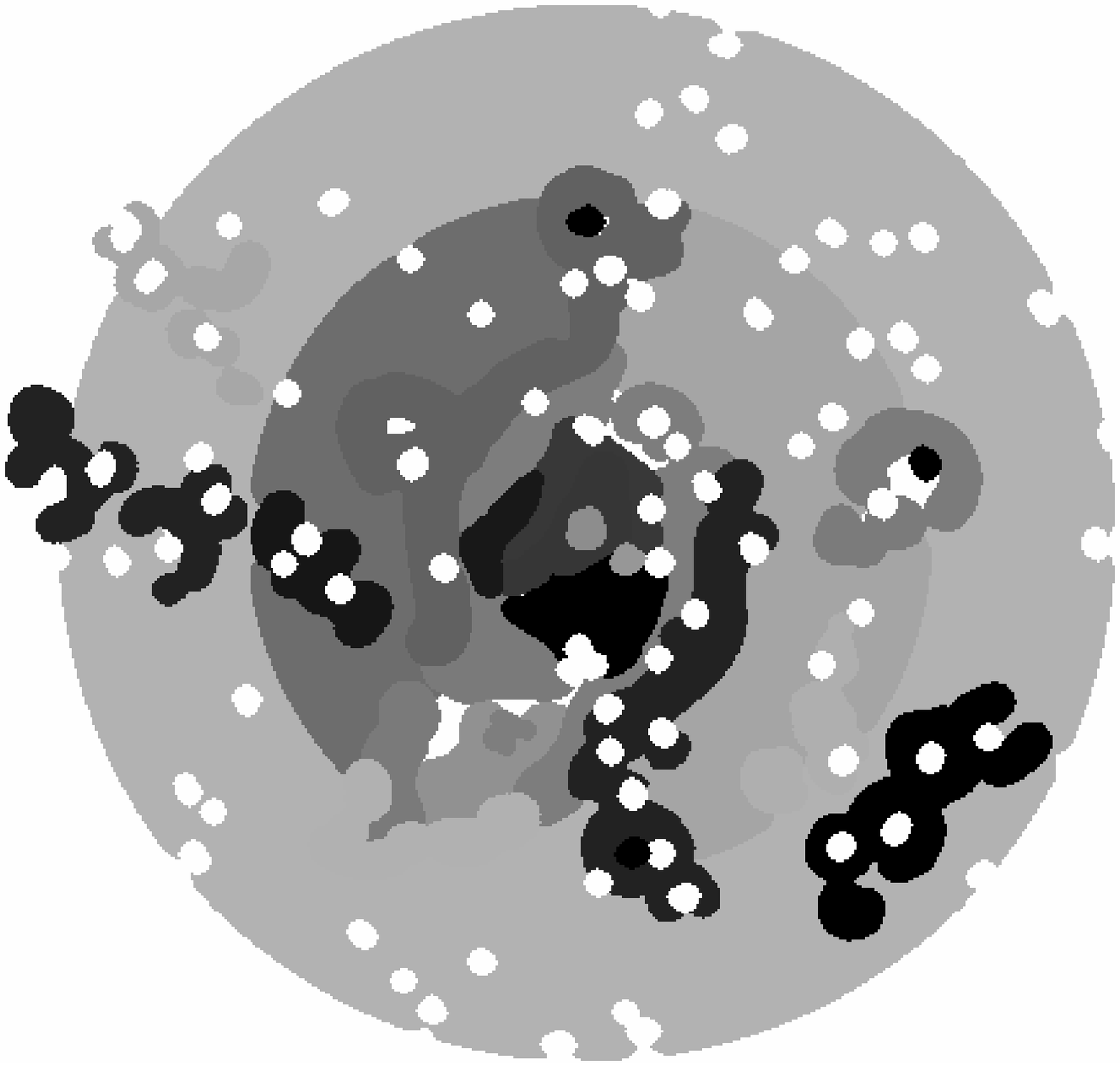}}\\ \figcaption{A
sample extraction template for HCG 97. The spectra in each region are
analyzed independently. White circles represent excised unrelated
sources. In this particular image, darker colors represent higher
emission-weighted plasma temperatures.\label{fig:region}}\\
\resizebox{3.5in}{!}{\includegraphics{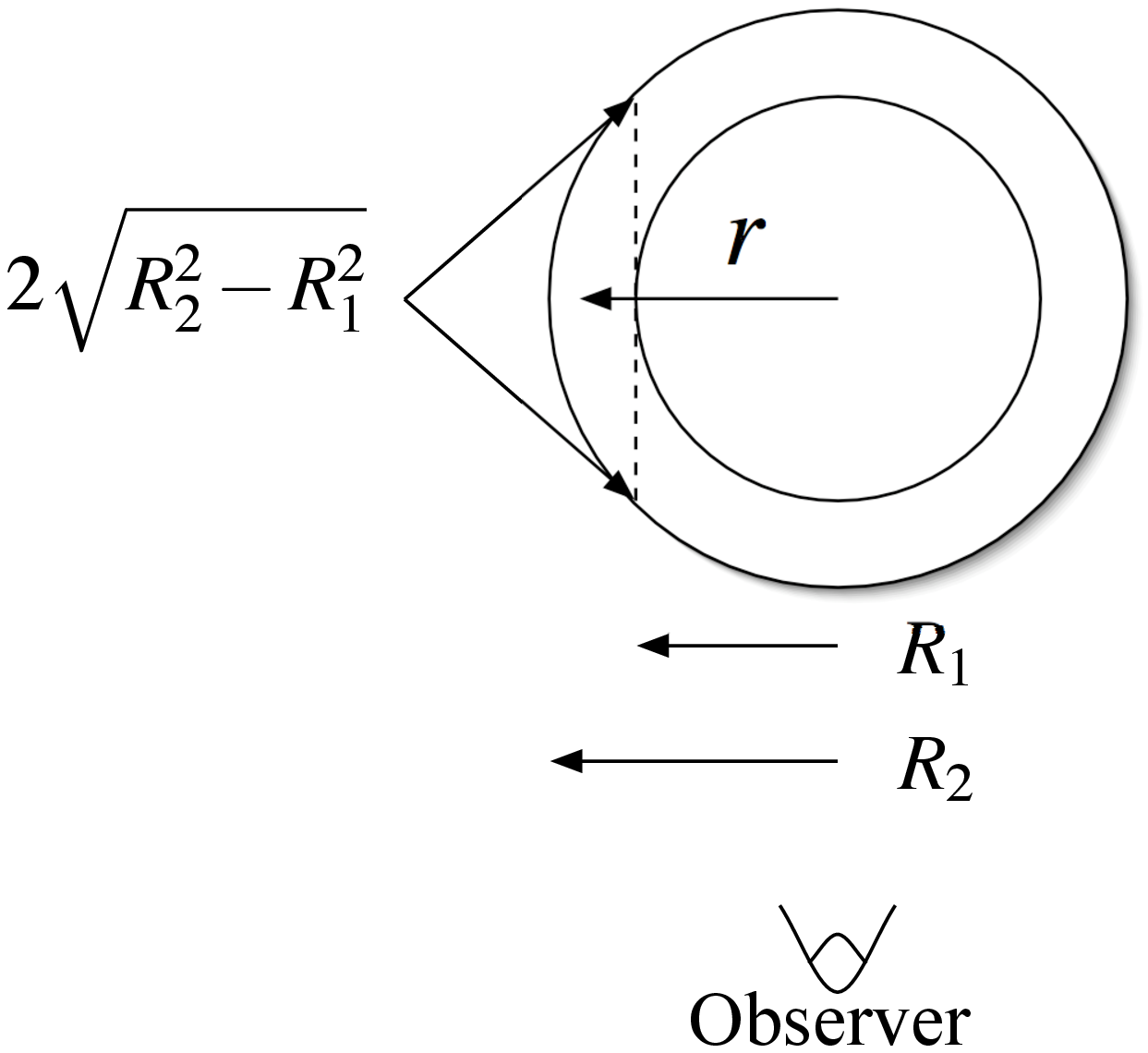}}\\
\figcaption{Simplification of the group geometry. The emitting region
is to the left of the dashed line, and has a minimum and maximum
projected distance $R_1$ and $R_2$, respectively, from the group
center.  The distance along the line of sight is then
$2\sqrt{R_2^2-R_1^2}.$
\label{fig:geometry}}
\end{tabular}

\begin{figure*}
\resizebox{7in}{!}{\includegraphics{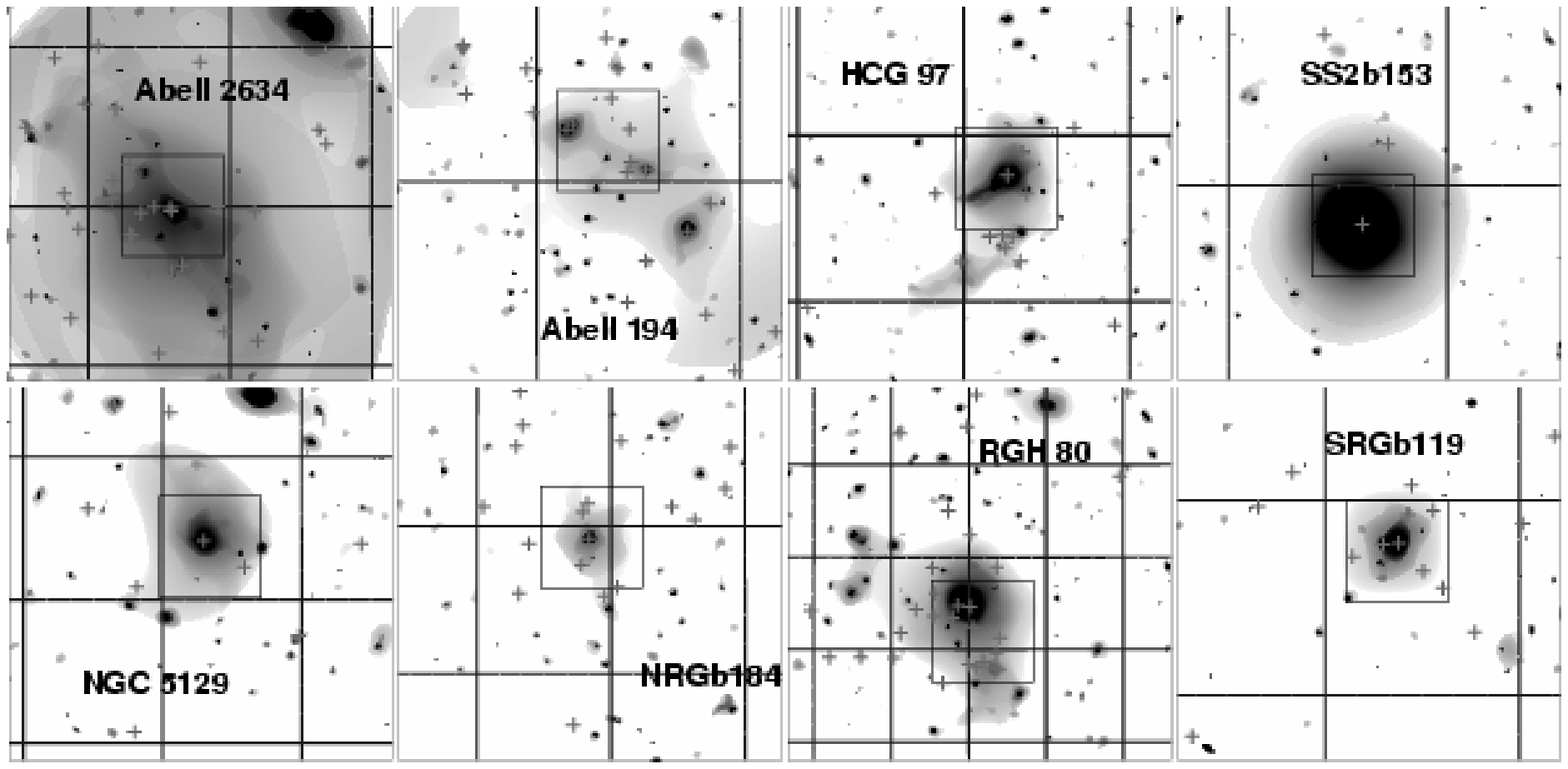}}
\resizebox{7in}{!}{\includegraphics{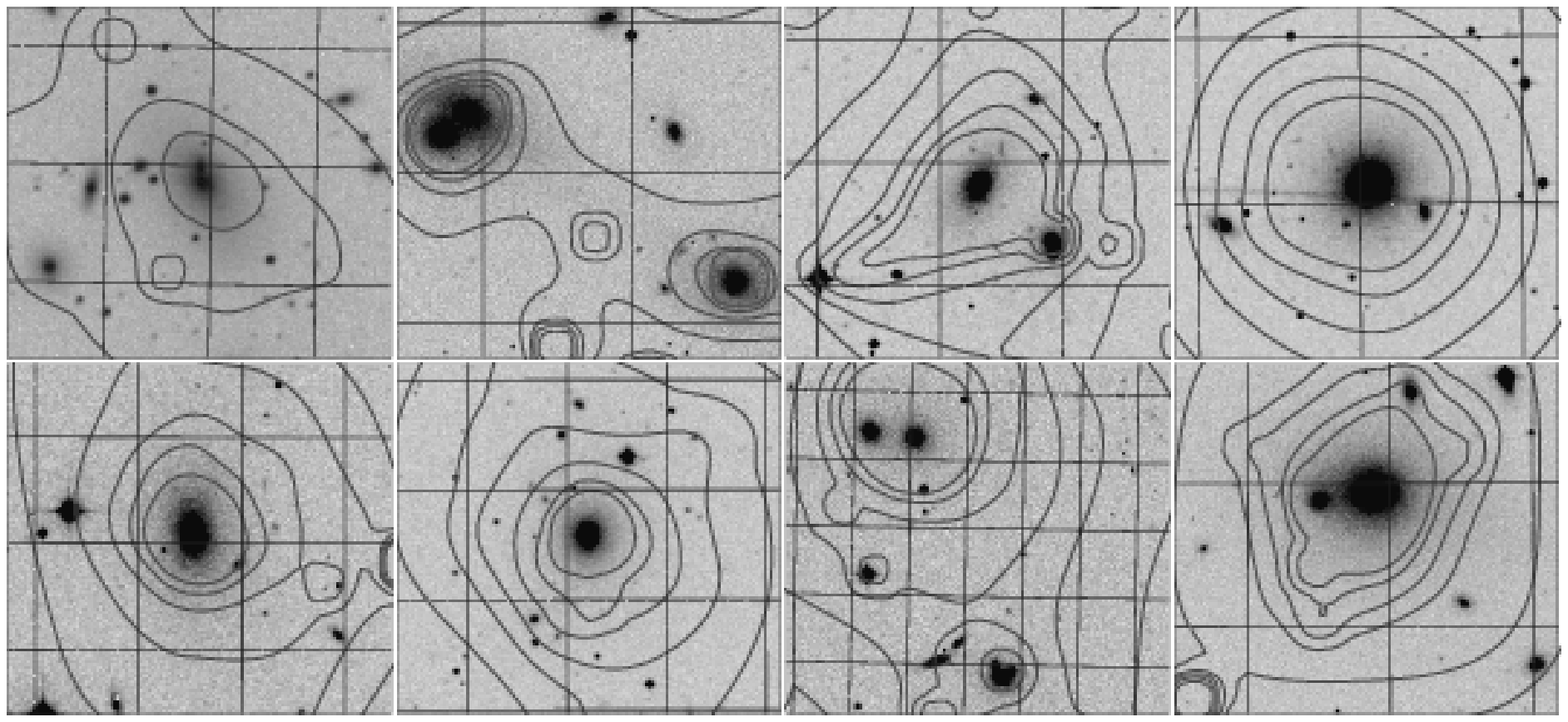}} \figcaption{X-ray and
optical images of the eight groups in our sample. (\emph{Top}) Wavelet
decomposition of the X-ray surface brightness in the 0.5-2.0 keV
energy band. Each image is $19\arcmin \times 19\arcmin$; the grid
cells are 200 kpc on a side. The small boxes show the area
enlarged below.  (\emph{Bottom}) Palomar Observatory Sky Survey images
of the central region of each group, with X-ray surface brightness
contours superimposed from the above images.  Each image is
$5.2\arcmin \times 5.2 \arcmin$ in size; the grid cells are $50$ kpc
on a side.
\label{fig:xrayopt}}
\end{figure*}

\begin{figure*}
\resizebox{7in}{!}{\includegraphics{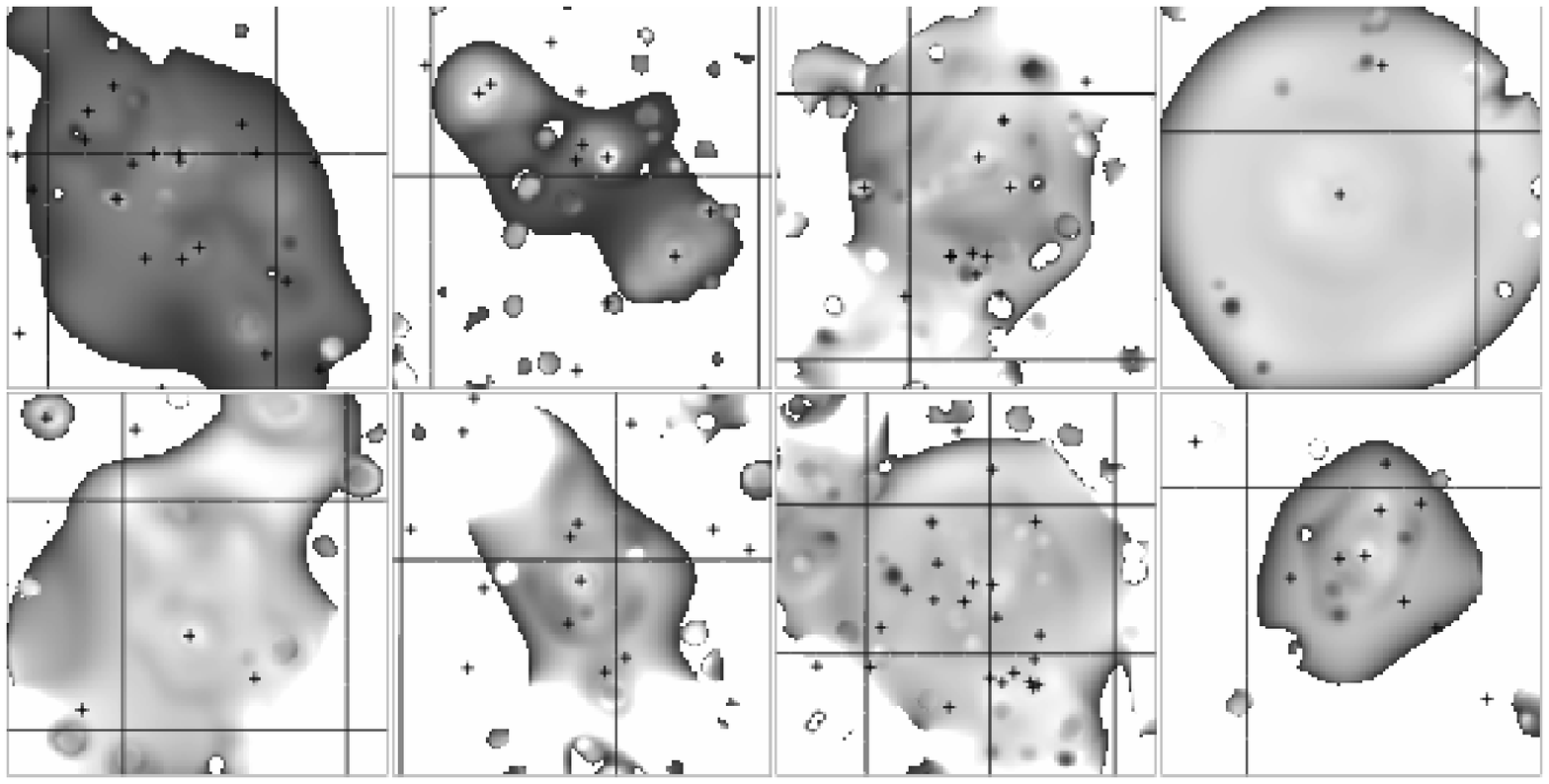}}
\resizebox{7in}{!}{\includegraphics{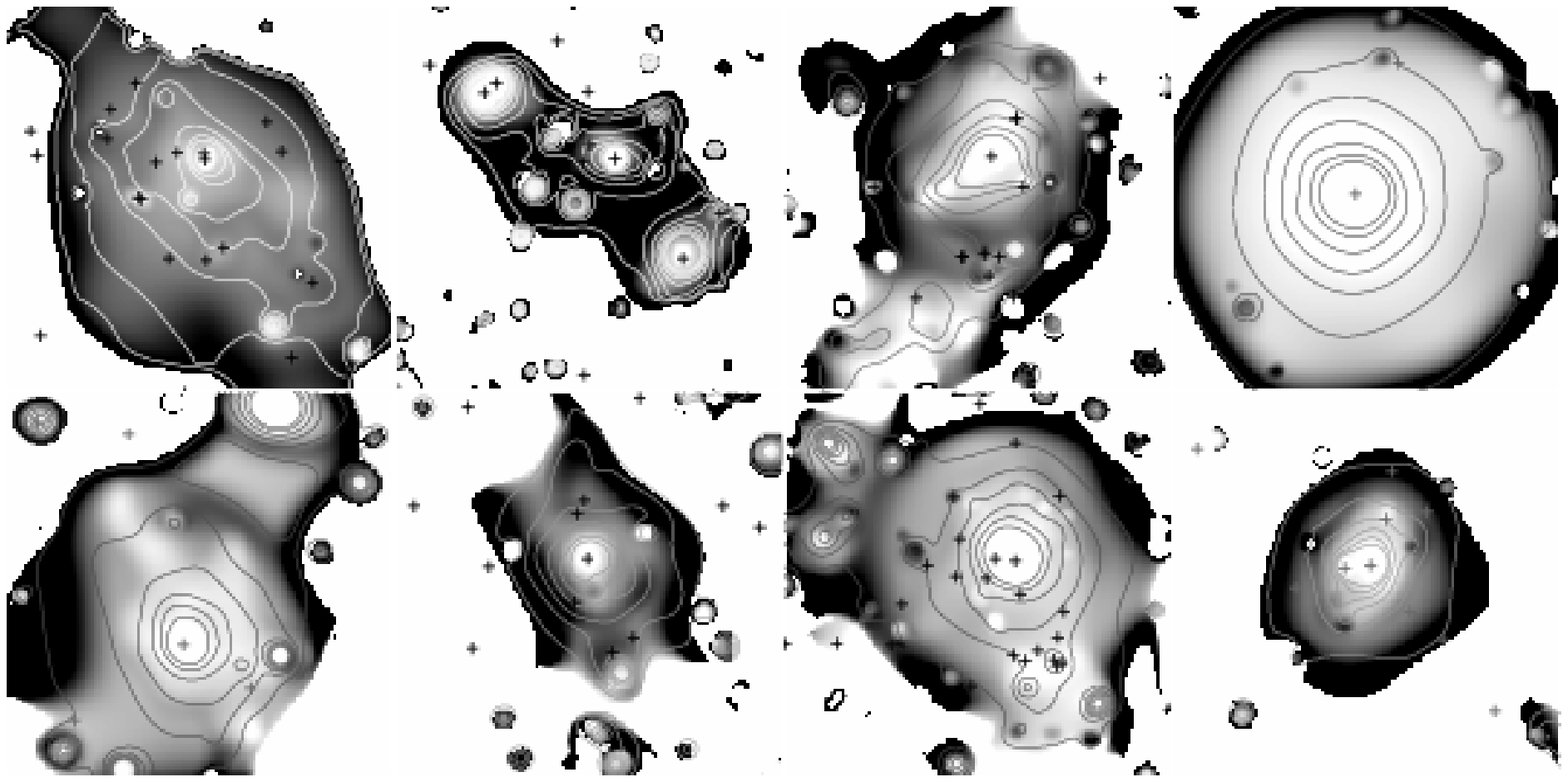}} \figcaption{(\emph{Top})
Emission-weighted temperature distribution of the sample from wavelet
analysis. The range in temperatures shown is 0.5 keV (white) to 2.3
keV (black), with a linear scale. Member galaxies are shown as
crosses. Each grid cell is 200 kpc on a side. (\emph{Bottom})
Distribution of the entropy ``integrated'' along the sight, estimated
from wavelet analysis. The range in integrated entropy shown is 10
(white) to 110 (black) in arbitrary units, with a linear scale.  The
contours show lines of constant gas pressure ``integrated'' along the
line of sight; the innermost contours always show higher pressure,
with each additional contour showing a decrement in pressure from the
previous one of a factor of 1.5.
\label{fig:snkt}}
\end{figure*}

Figures~\ref{fig:xrayopt} and~\ref{fig:snkt} show the gas density,
temperature, entropy, and pressure distribution generated by the
wavelet decomposition technique. The locations of the member galaxies
also appear in these figures. Table~\ref{tbl:xprops} lists the basic
physical properties of the X-ray emitting medium.

\section{Entropy and Pressure Profiles}

\label{sec:profiles}

The distribution of entropy in systems of galaxies has been used to
argue for the injection of ``extra energy'' into the intragroup medium
during the formation process. Several authors \nocite{Ponman99,Lloyd00}({Ponman} {et~al.} 1999; {Lloyd-Davies}, {Ponman}, \&  {Cannon} 2000)
have argued that in addition to gravitational collapse, processes such
as supernova explosions make a contribution to the thermodynamic state
of the gas. This contribution takes the form of a ``floor'' in the
inner entropy distribution of the intragroup medium as reported in
these studies. In other words, gas with higher entropy than expected
from pure gravitational collapse appears to be pooled at the center of
groups observed by the ROSAT satellite. \nocite{Ponman03}{Ponman}, {Sanderson}, \&  {Finoguenov} (2003) use a
larger sample of 66 systems observed by ASCA to study this
effect in greater detail. They find evidence of an entropy floor in
these observations, with the notable exception that in lower
temperature (less massive) systems the effect is much less pronounced
than in higher temperature (more massive) ones. The entropy profiles
of the less massive systems are better described by a simple power
law.

Further complicating the picture, \nocite{Finoguenov02}{Finoguenov} {et~al.} (2002), using
ASCA, found that gas at $r_{500}$ (the radius within which the
mean matter density is 501 times the critical density of the universe)
also exhibits excess entropy relative to the pure gravitational
value. This outer entropy excess seems consistent with models in which
shock heating and galactic winds are also a major contributor to the
groups' dynamical evolution \nocite{DosSantos02,Voit03}({Dos Santos} \& {Dor{\' e}} 2002; {Voit} \& {Ponman} 2003).  In addition,
\nocite{Ponman03}{Ponman} {et~al.} (2003) found that the entropy profiles, once scaled to
$T^{0.65}$, are in agreement with each other, hinting at the possible
universality of the entropy enrichment phenomenon. Some XMM-Newton
observations are consistent with this result \nocite{Pratt02}({Pratt} \& {Arnaud} 2002). Next we
revisit all of these results.

\subsection{Calculation of the Profiles}

\begin{figure*}
\resizebox{6.5in}{!}{\includegraphics{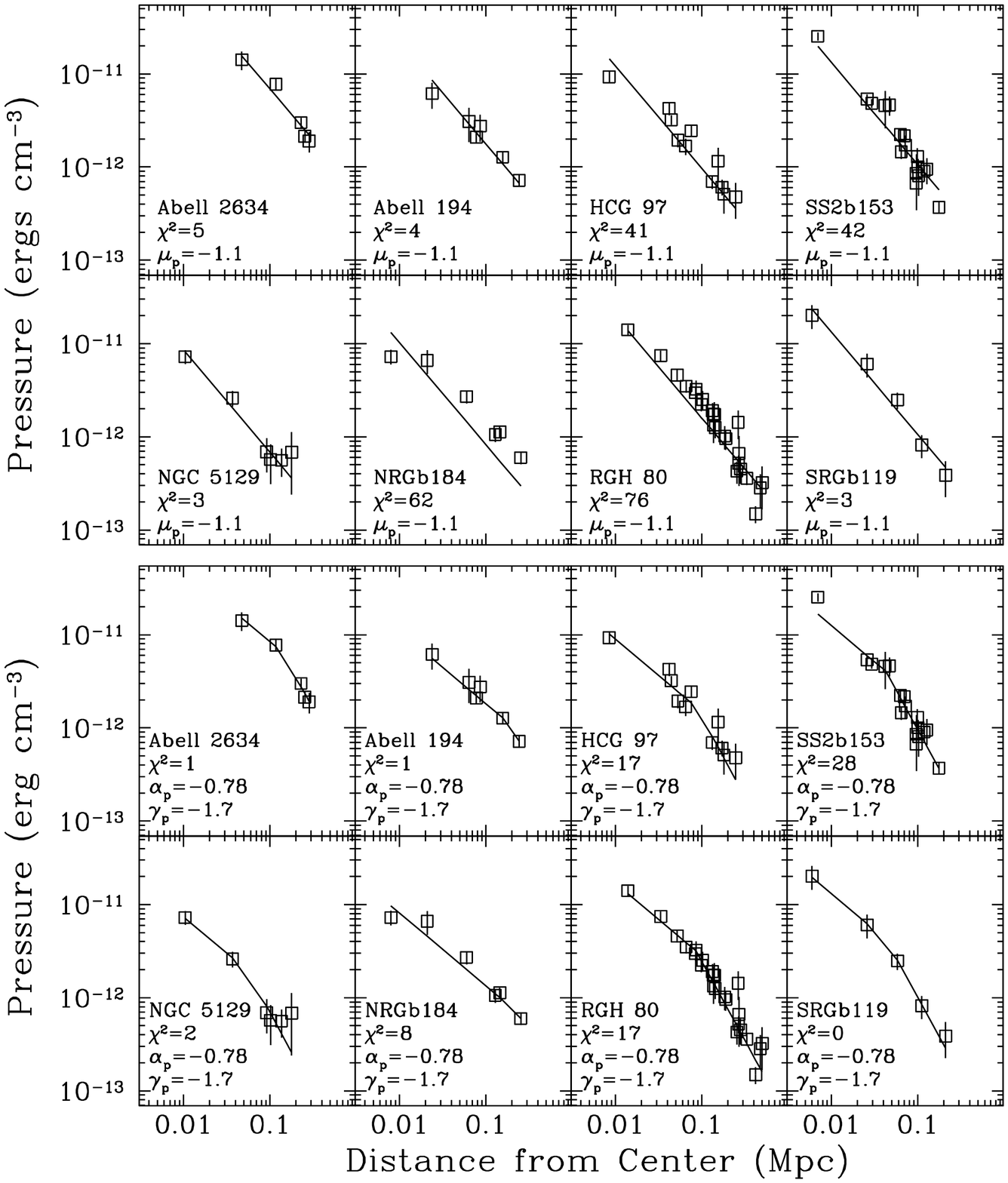}}
\figcaption{Pressure distributions for the eight groups in our
sample. (\emph{top}) Single-power law fit to the entire sample, with
the slope contrained to be same for each group; the reduced $\chi^2$
is 3.32. (\emph{bottom}) Broken power law fit; the reduced $\chi^2$ is
1.18.
\label{fig:pprof}}
\end{figure*}

\begin{figure*}
\resizebox{6.5in}{!}{\includegraphics{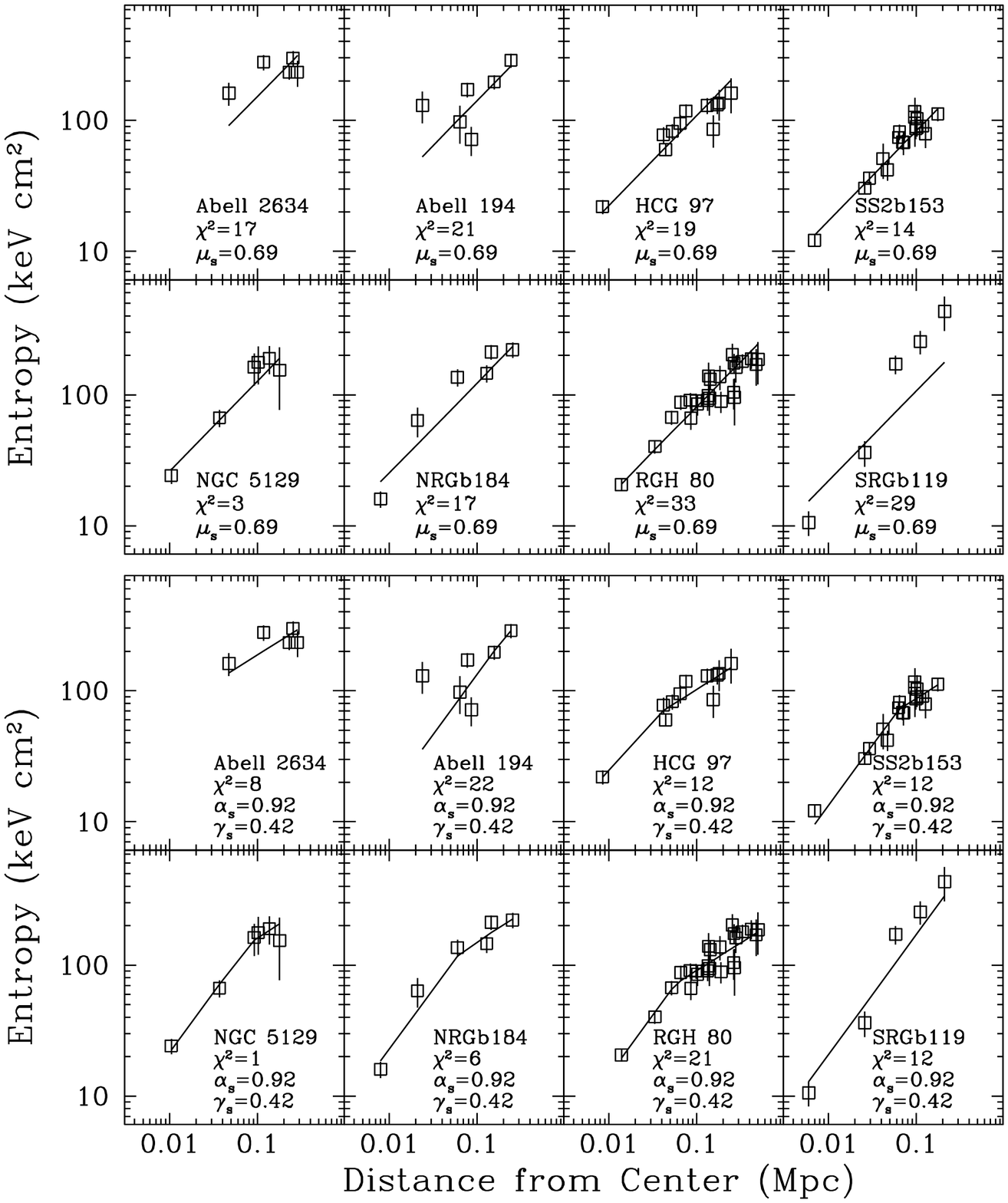}}
\figcaption{Entropy distributions for the eight groups in our
sample. (\emph{top}) Single-power law fit to the entire sample, with
the slope contrained to be same for each group; the reduced $\chi^2$
is 2.42. (\emph{bottom}) Broken power law fit; the reduced $\chi^2$ is
1.50.
\label{fig:sprof}}
\end{figure*}

Figures~\ref{fig:pprof} and \ref{fig:sprof} show the pressure and
entropy as a function of true (as opposed to projected) distance from
the center for our sample. The true distances were estimated not via
deprojection but using the method described in \S\ref{sec:xmm}
above. We define the gas entropy similarly to the earlier discussions
of the intragroup medium \nocite{Ponman03}({Ponman} {et~al.} 2003):
\begin{equation}
S \equiv T n^{-\frac{2}{3}},
\label{eq:entropy}
\end{equation}
where $n$ is the particle number density and $T$ is the temperature
in keV. As a result, $S$ has dimensions of keV cm$^2$. The
traditional, thermodynamical entropy is then $\log{S} + C$, where $C$
is an arbitrary constant.

We use the spectral fits discussed in \S\ref{sec:xmm} to plot the
entropy as a function of distance from the cluster center. The fits
are carried out in each region with the abundance, temperature, and
gas density as free parameters, and with the hydrogen column fixed to
the Galactic value from 21cm surveys \nocite{Dickey90}({Dickey} \& {Lockman} 1990). The results are
then used to calculate the entropy via equation (\ref{eq:entropy}). We
add an additional 10\% uncertainty to the entropy and pressure
calculated in this manner to account for inaccuracies involved in
our method of estimating distances from the center (\S\ref{sec:xmm});
the volume estimates are more uncertain than the distances.

Because the entropy and pressure profiles in the groups appear
similar, we test whether simple empirical models can describe their
shapes, and whether the same entropy and pressure profiles can
describe all the groups in our sample. Rather than scaling the data
by other derived values such as the temperature and $r_{500}$, we 
perform multidimensional fits (in up to 18 dimensions) as required
to compare the data with the self-similar models.

\subsection{Single Power Laws}

To begin, we test whether the entropy and pressure profiles can be
described by a single power law. From a theoretical point of view,
such a profile seems unlikely, because N-body simulations
\nocite{NFW,Ghigna00}({Navarro}, {Frenk}, \& {White} 1997; {Ghigna} {et~al.} 2000) suggest that the mass distribution in dark haloes
is not well fit by a single power law. The pressure and the
entropy distrbution depend on the temperature, and the temperature
profile in turn depends on the overall matter density distribution.
Thus it would be surprising if the resulting entropy and pressure were
scale-free. However, noise may well make the observations consistent
with a single power law, and it is important to test for this
possibility before considering more complicated models.

We wish to discover whether the pressure and entropy distributions in
our sample of groups are self-similar.  Therefore, in fitting single
power laws to the data, we constrain the slope of the power laws to be
the same for all eight groups: $ S(r) \propto r^{\mu_s}$ for the
entropy, and $P(r) \propto r^{\mu_p}$ for the pressure. Each fit has 9
free parameters: one slope and eight normalizations. We minimize the
$\chi^2$ statistic via the built-in implementation of simulated
annealing \nocite{NR}({Press} {et~al.} 1992) in the Sherpa data analysis
package\footnote{Sherpa was used for fitting the radial profiles only;
details on the software may be found at
\url{http://asc.harvard.edu/sherpa/}}. Simulated annealing is a
compromise between downhill minimization (which is fast but may not
always find a global minimum) and monte-carlo minimization (which is
slow but has a greater likelihood of finding a global minimum).

The results appear in the top halves of Figures \ref{fig:pprof} and
\ref{fig:sprof}. The fits are not acceptable, yielding reduced
$\chi^2$ of 2.42 and 3.32 for the entropy and pressure data,
respectively. We conclude that neither the entropy nor the pressure
profiles are consistent with a single universal power law.

We can proceed either by treating the groups as individual entities---
fitting independent single power laws to each system---or by
considering more complicated universal models. The first approach
yields an acceptable fit for the entropy profiles of Abell 2634
($\mu_s = 0.21 \pm 0.10$), NGC 5129 ($\mu_s = 0.79 \pm 0.08$), HCG 97
($\mu_s = 0.59 \pm 0.04$), RGH 80 ($\mu_s = 0.62 \pm 0.03$), SS2b153,
($\mu_s = 0.72 \pm 0.04$), and SRGb119 ($\mu_s = 1.1 \pm 0.1$); Abell
194 and NRGb184 do not yield acceptable fits. For the pressure
profiles, single power laws yield a good fit to Abell 2634 ($\mu_p =
-1.2 \pm 0.1$), NGC 5129 ($\mu_p = -1.0 \pm 0.1$), Abell 194 ($\mu_p =
-0.92 \pm 0.11$), NRGb184 ($\mu_p = -0.73 \pm 0.05$), SS2b153 ($\mu_p
= 1.26 \pm 0.04$), SRGb119 ($\mu_p = 1.1 \pm 0.09$); HCG 97 and RGH 80
are not consistent with a single power law.  Interpreting these
individual slopes is by no means straightforward, especially in light
of the fact that we may not be sufficiently sampling the central
regions of the more distant systems. It is far more instructive to
consider whether a more detailed universal profile can describe all
the data.

\subsection{Broken Power Laws}

We test the self-similarity of the entropy and pressure profiles
by fitting broken power laws of the form
\begin{equation}
P(r) =  P_0 
    \cases{ (r/100\mr{\ kpc})^{\alpha_p}  &  $r < r_p$ \cr
   (r_p/100\mr{kpc})^{\alpha_p-\gamma_p} (r/100\mr{\ kpc})^{\gamma_p} & $r > r_p$
   }
\end{equation}
Similarly, we fit entropy profiles of the form
\begin{equation}
S(r) =  S_0 
   \cases{  (r/100\mr{kpc})^{\alpha_s}  &  $r < r_s$ \cr
   (r_s/100\mr{kpc})^{\alpha_s-\gamma_s} (r/100\mr{kpc})^{\gamma_s} & $r > r_s$
   }
\end{equation}

For these self-similar models we conduct two separate analyses: a
pressure fit and an entropy fit. The pressure fit has 18 free
parameters: the slopes $\alpha_p$ and $\gamma_p$ common to all the
groups, and eight different pairs ($r_p,P_0)$ for each
group. Similarly, the entropy fit has its own set of 18 parameters. 

Minimizing $\chi^2$ using the same technique as in the previous
section yields good fits; these are shown in the bottom half of
Figures \ref{fig:pprof} and \ref{fig:sprof}. In Table
\ref{tbl:results}, we estimate the goodness-of-fit (the probability
of observing $\chi^2 > \chi^2_\mr{min}$ if the model is correct) as
the definite integral of the $\chi^2$ distribution,
$q=\Gamma(\nu/2,\chi^2/2)/\Gamma(\nu/2)$, where $\Gamma(a,x)$ is the
incomplete gamma function, $\Gamma(a) = \Gamma(a,0)$ is the gamma
function, and $\nu$ is the number of degrees of freedom (i.e., the
number of data points minus the number of free parameters minus one).

After finding the minimum, we calculate errors on the best-fit
parameters by using likelihood ratio tests \nocite{Lupton}({Lupton} 1993). This method
fully takes the correlations among the parameters into account. The
test involves repeatedly comparing the overall minimum $\chi^2$ with
$\chi^2$ minimized assuming many different fixed values for the
parameter in question.  The set of differences between the global
minimum and the constrained minima has a $\chi^2$ distribution with
one degree of freedom asymptotically, and this fact may be used to
estimate confidence intervals. For each parameter we re-minimize
$\chi^2$ 100 times around the best-fit value; the error estimation
procedure thus involved 1800 re-minimizations in all.

\begin{figure*}
\resizebox{7in}{!}{\includegraphics{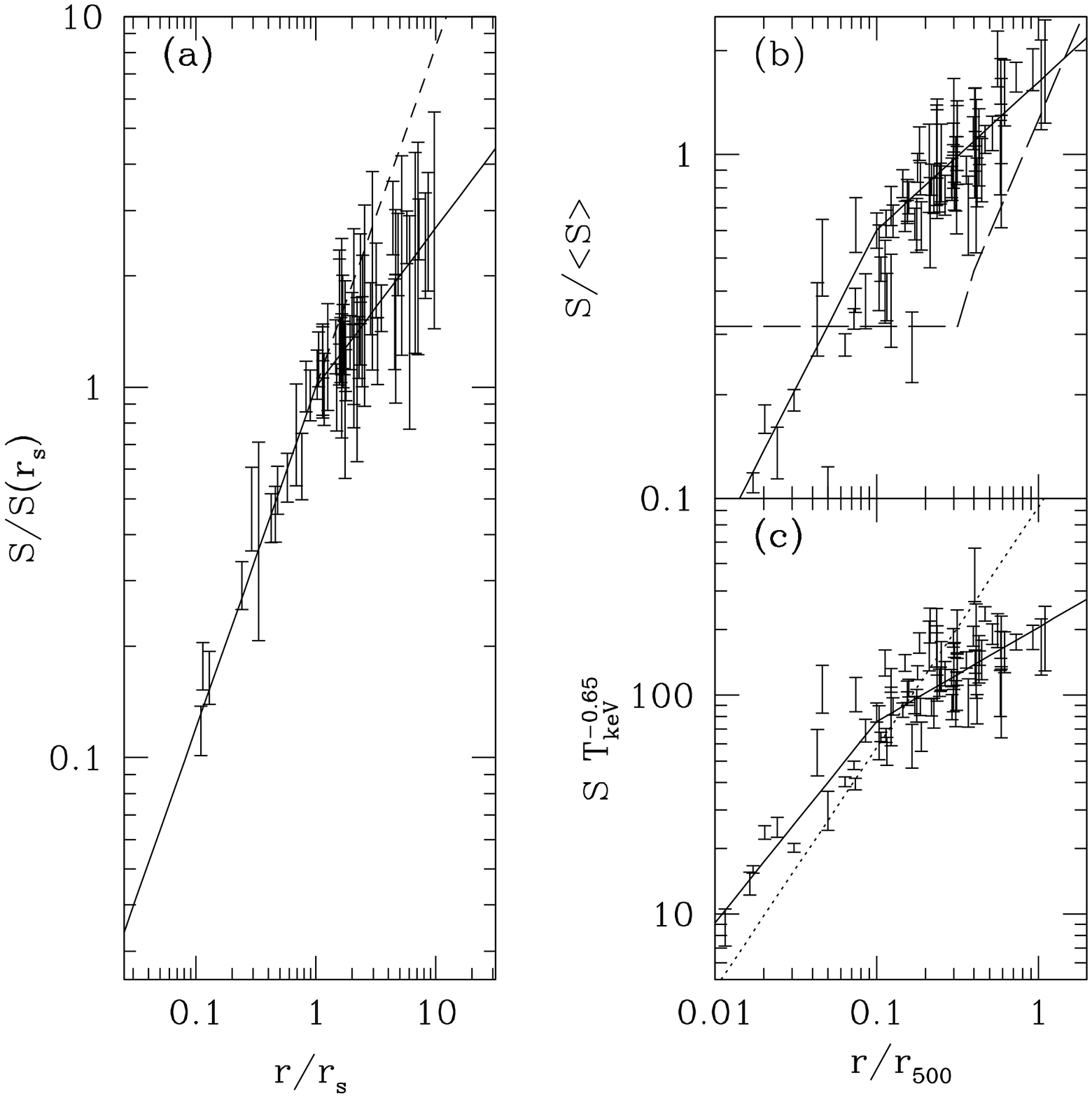}} \figcaption{The
entropy profiles combined using (\emph{a}) our fit results, but
leaving out groups where $S(r)$ is consistent with a single power
law; (\emph{b}) all groups, with the radii scaled by $r_{500}$ and
the entropies scaled by the mean entropy $\langle S \rangle$ within
$r_{500}$; and (\emph{c}) all groups with the radii scaled by
$r_{500}$ and the entropies combined according to the scaling relation
derived by \protect\nocite{Ponman03}{Ponman} {et~al.} (2003).  The solid curve in all three
figures shows the broken power law derived from our analysis,
$\alpha_s = 0.92$ and $\gamma_s = 0.42$, with the breaking radius and
normalization adjusted to fit the data in (\emph{b}) and
(\emph{c}). The short-dashed line in (\emph{a}) shows the continuation
of the $\alpha_s = 0.92$ power law. The long-dashed curve in
(\emph{b}) shows the prediction of the \protect\nocite{Tozzi01}{Tozzi} \& {Norman} (2001)
preheating model. The dotted curve in (\emph{c}) shows the (rejected)
best-fit power law with slope fixed at 1.1.
\label{fig:combine}}
\end{figure*}

\subsection{Results}

We find that a broken power law provides a good description of the
combined pressure profiles of the eight groups (see Table
\ref{tbl:results} and Figure~\ref{fig:pprof}).  The pressure broken
power law has a shallow slope \fitalphap\ near the group center,
steepening to \fitgammap\ at the outer edges; the goodness-of-fit
is an acceptable $q=0.15$. The agreement is not perfect: NRGb184 is
entirely consistent with a single, rather than a broken power law, and
SS2b153 exhibits a somewhat steeper inner pressure profile than the
other seven groups. Otherwise, the notion that the groups are
self-similar systems differing only in a radius and absolute pressure
scaling is supported by the XMM-Newton data.

The entropy profiles (Figure~\ref{fig:sprof}), however, make the
picture more complicated.  The broken power law fit to the combined
data has fit quality $q=0.0095$ ; the model is acceptable at
$3\sigma$. Excluding the binary cluster Abell 194 would bring the
goodness-of-fit to a wholly acceptable $q=0.136$, and would not affect
the best-fit slopes substantially. The inner power law slope,
\fitalphas, is steeper than the outer entropy slope \fitgammas,
meaning there is no evidence for an constant-entropy ``floor'' as
close as 10 kpc from the center of each group. Rather, in contrast to
previous observations with ROSAT and ASCA, the entropy
profile shows a tendency to steepen towards the center of each group.
There is an entropy deficit at larger radii relative to the
extrapolation one would obtain from the inner power law, a new result
not seen in data from the older X-ray observatories.

It is instructive to compare our data with theoretical models.
\nocite{Tozzi01}{Tozzi} \& {Norman} (2001) make detailed predictions for the entropy profiles of
groups and clusters of galaxies in two different scenarios: with
preheating (where a significant initial entropy excess exists in the
systems at $t \approx 1$ Gyr), and without preheating (where the gas
dynamics is chiefly determined by shocks and gravitational
heating). In the first scenario, they find that more massive clusters
develop isentropic cores within $\sim 0.05 r_{200}$, and that less
massive groups develop relatively larger cores (within $\sim 0.5
r_{200}$). By contrast, in the scenario without preheating, groups and
clusters both tend to develop power law entropy distributions, with
$S(r) \propto r^{1.1}$.

To test these scenarios, we combine our entropy profiles in several
ways (Figure~\ref{fig:combine}).  In (\emph{a}), we show all the
groups for which the preferred fit is a broken power law---that is,
the groups for which $r_s$ is neither a lower nor an upper limit. Here
the radii are scaled by $r_s$, and the entropies by $S(r_s)$. As might
be expected, this combination of profiles exhibits little scatter
around the best-fit line, because the highest contributors to
$\chi^2$---Abell 194 and SRGb119---are not included. In (\emph{b}) we
show the profiles for all eight groups, scaled by $r_{500}$ and
$\langle S \rangle$, the mean entropy within $r_{500}$ as listed in
Table~\ref{tbl:xprops}. Despite the difference in scaling, the data
are still well described by our best-fit broken power law with fixed
slopes $\alpha_s=0.92$ and $\gamma_s=0.42$---in fact, the fit quality
is higher, because the additional uncertainty introduced through
division by $\langle S \rangle$ decreases the reduced $\chi^2$.  We
compare these profiles with the predictions of the \nocite{Tozzi01}{Tozzi} \& {Norman} (2001)
preheating model, where we assume that $\langle S \rangle =
S_\infty/2$, where $S_\infty$ is the entropy of the last accreted gas
shell. The model is rejected, because it predicts a large isentropic
core that we do not observe and an outer profile that rises more
quickly than ours. Varying $\langle S \rangle/S_\infty$ arbitrarily in
either direction does not change this result. Finally, in (\emph{c})
we show the eight groups again, with the radii scaled by $r_{500}$,
but with the entropies scaled by $(T/\mr{keV})^{0.65}$, a relation
\nocite{Ponman03}{Ponman} {et~al.} (2003) derive for their sample of 66 virialized groups and
clusters. This scaling agrees with our $\alpha_s=0.92$ and
$\gamma_s=0.42$ fits as well, but is clearly inconsistent with both
the preheated and non-preheated \nocite{Tozzi01}{Tozzi} \& {Norman} (2001) models---both models
are rejected with $q<10^{-6}$.

Thus there are two key differences between our results and the
\nocite{Tozzi01}{Tozzi} \& {Norman} (2001) models: (1) we do not observe any isentropic cores, and
(2) we find an entropy decrement at radii $\gtrsim 0.5 r_{500}$ with
respect to the extrapolation from data at radii $\lesssim 0.5
r_{500}$. Neither of these results is unique to our sample: both
effects are also clear in the high quality Chandra observation of the
NGC 1550 group \nocite{Sun03}({Sun} {et~al.} 2003); the lack of an isentropic core is also
clear in the Chandra and XMM-Newton observations of ESO 306170
\nocite{Sun04}({Sun} {et~al.} 2004) and in a few of the groups studied by
\nocite{Ponman03}{Ponman} {et~al.} (2003). These studies employ deprojection techniques very
different from our mask-based approach; thus it is unlikely that we
are witnessing a peculiar effect due to the instrument or due to our
data reduction technique.

In the inner regions, the disagreement with older studies may be due
to the improved spatial resolution of Chandra and XMM-Newton. In
studies such as \nocite{Ponman03}{Ponman} {et~al.} (2003), the distribution of the gas in groups
is assumed to follow a single $\beta$-model distribution, with gas
density $\rho_g \propto (1+r^2/r_c^2)^{-3 \beta/2}$
\nocite{Sanderson03}({Sanderson} {et~al.} 2003). Higher resolution data indicate, however, that
the gas density rises more steeply towards the center due to the
influence of the dominant central galaxy
\nocite{Helsdon00,Lewis03,Buote03,Sun03,Sun04}({Helsdon} \& {Ponman} 2000; {Lewis}, {Buote}, \& {Stocke} 2003; {Buote} {et~al.} 2003; {Sun} {et~al.} 2003, 2004), with $n$ better
described by a broken power law or by two superposed
$\beta$-models. In this scenario, as $r \rightarrow 0$, the entropy $S
\propto T / n^{2/3}$ declines more rapidly than it does for an
equivalent single $\beta$ model, unless $r_c$ is very small.

The disagreement with the theoretical models predicting large
isentropic cores is likely related to the assumed equation of state of
the gas. In such models \nocite{Tozzi00,Tozzi01}({Tozzi}, {Scharf}, \& {Norman} 2000; {Tozzi} \& {Norman} 2001), the initial gas
density profile follows a polytropic distribution, $n \propto \
p^{1/\eta} \propto T^{1/(\eta-1)}$, with $\eta$ being the adiabatic
index.  Given this functional relationship, the temperature gradient
$d T /d r$ is inevitably negative everywhere, because the gas density
must decline with radius. This scenario, when combined with an initial
entropy excess (due to supernova-driven galactic winds, for example),
yields an isentropic core that grows as the group evolves.  However,
the temperature distribution in this preheating scenario is in strong
contrast to the data, where, at least within $0.3 r_{500}$, the
temperature typically exhibits a positive gradient, rising
monotonically from the center of groups as well as relaxed rich
clusters of galaxies, and declining or remaining constant only outside
$0.3 r_{500}$ \nocite{Pratt02,Zhang04}({Pratt} \& {Arnaud} 2002; {Zhang} {et~al.} 2004). Thus there is a fundamental
problem with the inner temperature structure predicted by the
polytropic preheating models.  In the era of ROSAT and ASCA, this
disagreement was sometimes explained with the argument that the inner
regions of clusters and groups contain cooling flows, which disrupt
the equilibrium physics, and which therefore need to be excised
\nocite{Markevitch98}({Markevitch} 1998) or modeled as two-temperature systems
\nocite{Ettori00}({Ettori} 2000). However, Chandra and XMM-Newton show that
hydrostatic models---in many cases including the region within the
cooling radius---can fit clusters with rising temperature profiles
\nocite{Schmidt01,Allen02,Ettori02,Buote03,Zhang04}({Schmidt}, {Allen}, \& {Fabian} 2001; {Allen}, {Schmidt}, \& {Fabian} 2002; {Ettori} {et~al.} 2002; {Buote} {et~al.} 2003; {Zhang} {et~al.} 2004). 

Therefore, rising temperature profiles must be a part of any
successful preheating model. The preheating models of
\nocite{Muanwong02}{Muanwong} {et~al.} (2002), \nocite{Tornatore03}{Tornatore} {et~al.} (2003), and \nocite{McCarthy04}{McCarthy} {et~al.} (2004), for
example, do away with the polytropic assumption, and therefore can
reproduce the positive inner temperature gradients observed in many
clusters.  These works show that when radiative cooling is also
included, the models with and without preheating produce nearly
indistinguishible radial entropy profiles. In all these calculations,
regardless of the treatment of cooling, the large isentropic cores
vanish.  It is interesting to note that in all cases the non-preheated
scenarios provide the most satistifactory description of the our
XMM-Newton data, predicting entropy distributions with slopes close to
our observed of \fitalphas\ within $0.3 r_{500}$. In contrast,
depending on the redshift at which the energy injection occurs, the
preheated models predict shallower inner slopes than we observe. Thus
the ``similarity break'' in the cluster luminosity-temperature
relation---one of the original motivations for the notion of
preheating---continues to elude self-consistent modeling.

The second chief disagreement---the entropy deficit at the outer radii
near $r_{500}$---is more difficult to explain. While it is possible
that substructure or superposition with background clusters could
cause the deviation from a single power law, we do observe the
decrement in groups that show no evidence of irregularity, such as
SS2b153, NRGb184 and NGC 1550 \nocite{Sun03}({Sun} {et~al.} 2003). An understanding of this
entropy deficit will likely require higher signal-to-noise data as
well as preheating models that more accurately reflect the temperature
distribution of the intragroup medium.

\resizebox{3.5in}{!}{\includegraphics{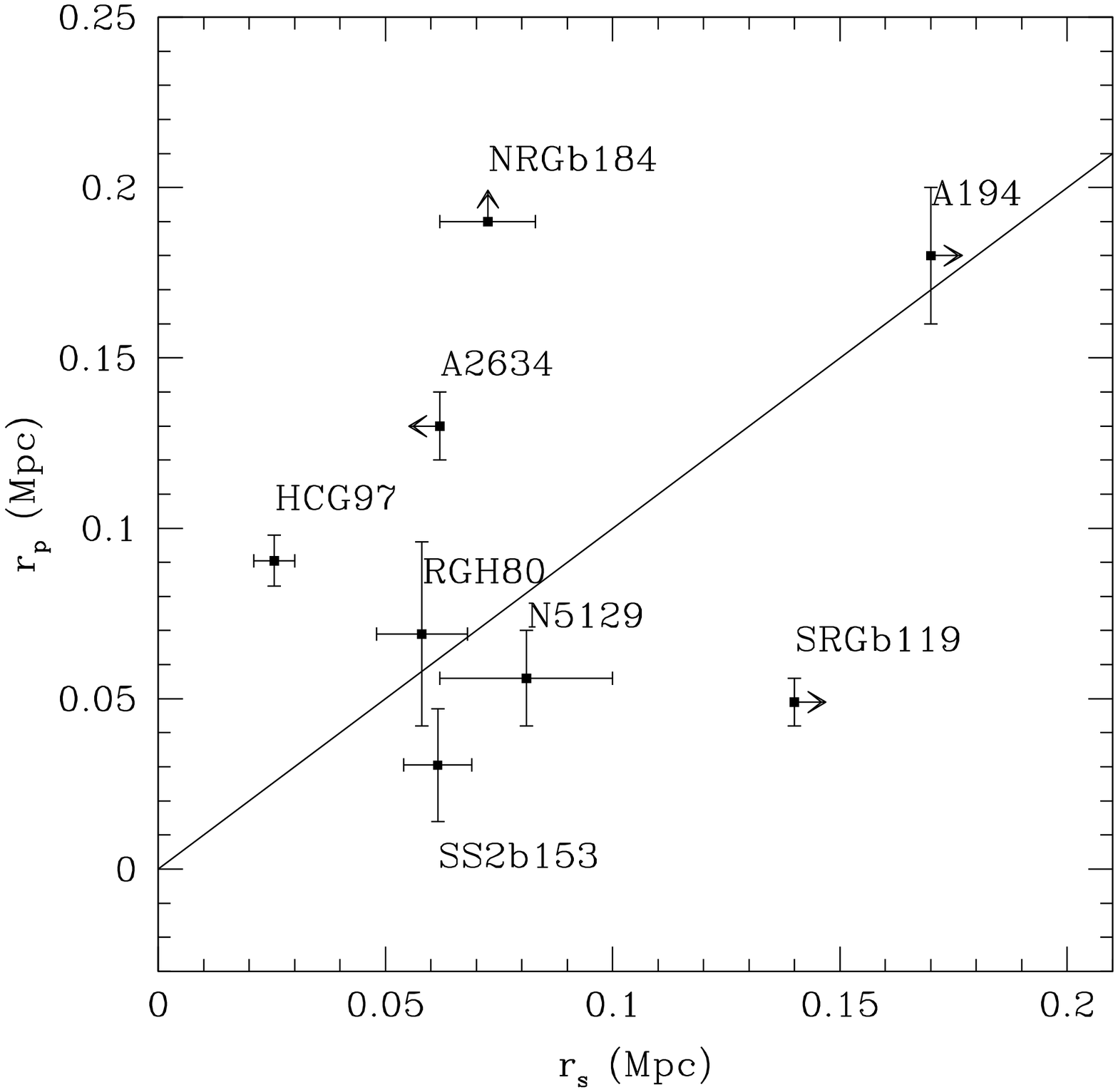}} \figcaption{The
pressure transition slope $r_p$ compared to the entropy transition
slope $r_s$. The solid straight line corresponds to $r_s = r_p$;
arrows indicate upper or lower limits. 
\label{fig:rprob}}
\vspace{0.5in}

\begin{figure*}
\resizebox{3.4in}{!}{\includegraphics{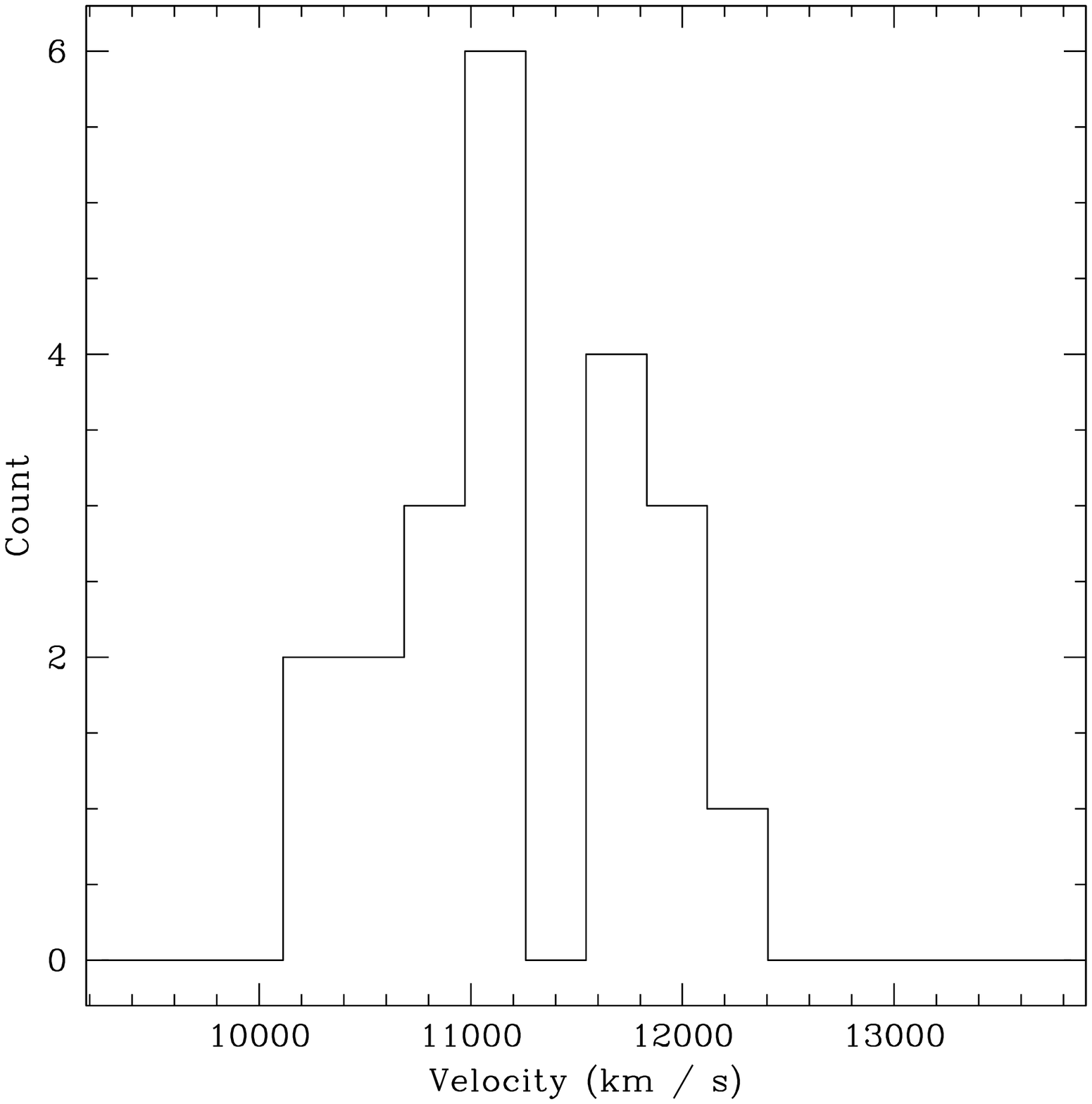}}
\resizebox{3.6in}{!}{\includegraphics{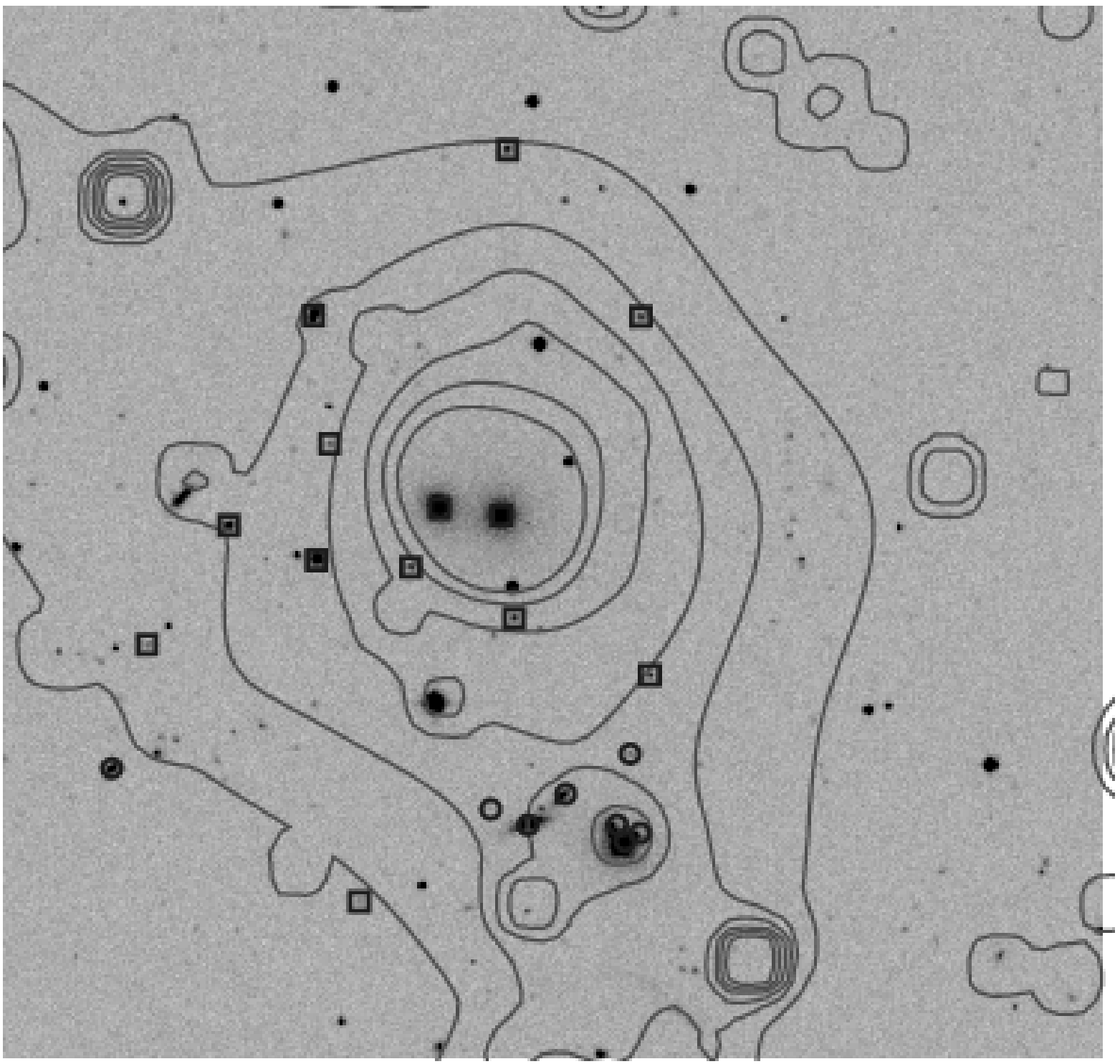}}
\figcaption{(\emph{Left}) Histogram of velocities for galaxies within
12\arcmin\ of the center of the RGH 80 group. There is a gap
of $490$ km s\m\ between the two peaks, comparable to the group's
overall velocity dispersion. (\emph{Right}) POSS image of RGH80,
showing galaxies belonging to the histogram's low velocity peak
(squares) and high velocity peak (circles). The circles are
concentrated around the SW X-ray emission peak. \label{fig:rgh80hist}}
\end{figure*}

\begin{figure*}
\resizebox{7in}{!}{\includegraphics{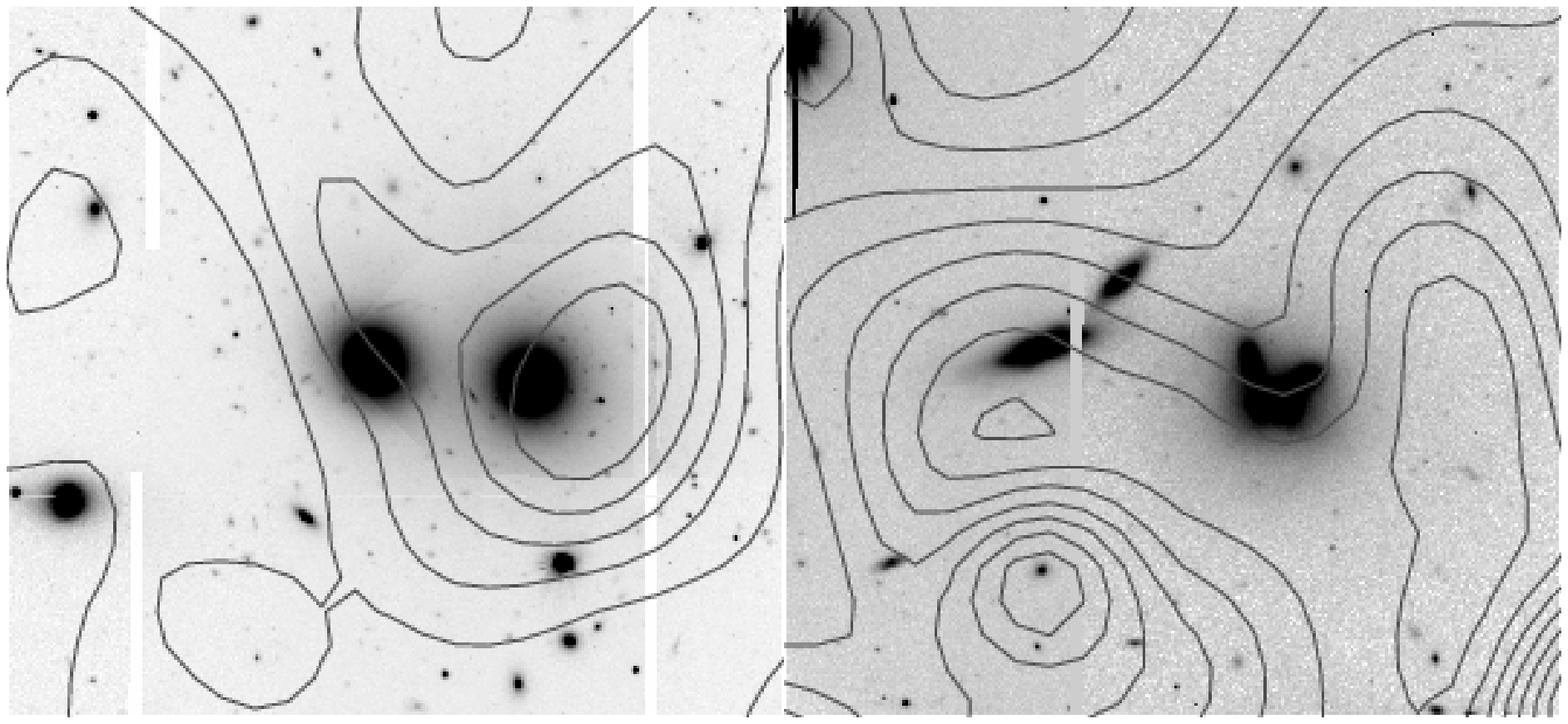}}
\figcaption{$R$-band Gemini GMOS acquisition images of RGH80, showing
the dominant X-ray peak (\emph{left}) and the subgroup to the SW
(\emph{right}). Also shown are lines of constant emission-weighted
temperature. In each case, the innermost contour shows a temperature
of 0.6 keV; each larger contour corresponds to an increment of 0.05
keV from the previous one.
 \label{fig:rgh80gemini}}
\end{figure*}

Returning to the self-similar fits, we briefly comment on the physical
significance of the independently derived transition radii $r_s$ and
$r_p$. Because these breaking radii are typically $\gg 10$ kpc, it is
unlikely that they are simply related to the effect of the potential
of the central galaxy on the gas distribution. Similarly, most of the
measured radii are larger than the size of a typical group-scale
cooling flow ($\approx 20$ kpc). Thus, it is more likely that the
transition radii are reflective of the properties of the gas
distribution throughout the group, rather than just the effect of a
central galaxy or cooling flow. In that case, given a perfectly
self-similar sample, one would expect a monotonic functional
relationship between $r_s$ and $r_p$: as the entropy transition radius
increases or decreases, the pressure radius should follow suit. The
simplest such scenario would be $r_s = r_p$, i.e., the entropy and the
pressure profiles both breaking at the same point. We examine this
possibility in Figure~\ref{fig:rprob}. Here we see that half the
sample does lie close to the $r_s = r_p$ relationship: RGH 80, NGC
5129, SS2b153, and Abell 194 are within 1 $\sigma$ of the line. Of
these, the first three appear relaxed in both their X-ray and optical
morphology (see \S\ref{sec:individual} below). Of systems that deviate
from the $r_s = r_p$ line, Abell 2634, SRGb119, and HCG 97 show hints
of irregularities that we discuss in \S\ref{sec:individual} below,
while NRGb184 appears relaxed. Thus the relationship between $r_s$ and
$r_p$ is a useful (if not foolproof) tool in evaluating the
equilibrium state of a system of galaxies.

Thus the XMM-Newton observations demonstrate that our sample of
groups has a self-similar azimuthally averaged pressure distribution.
The entropy profiles also appear similar. These entropy profiles have
shapes that challenge detailed theoretical models that include
preheating \nocite{Tozzi00,Tozzi01}({Tozzi} {et~al.} 2000; {Tozzi} \& {Norman} 2001). Furthermore, non-preheated models
are valid only in the $\emph{inner}$ regions---we find $S(r) \propto
r^{0.91\pm0.04}$, closely matching the the models' predicted $S(r)
\propto r^{1.1}$.  However, as our data approaches $r_{500}$, the
slope flattens, even in groups where the gas distribution appears
regular. Present preheating models do not account for the $\gamma_s
\approx 0.4$ slope we find near $r_{500}$.

\section{Properties of Individual Systems}
\label{sec:individual}

Before we discuss the properties of the individual systems, the
heterogeneous nature of the sample as a whole is worth pointing
out. No two systems in our sample resemble one another
morphologically. There is a spherical and relaxed-seeming group
(SS2b153), a \nocite{Hickson82}{Hickson} (1982) compact group merging with a sub-group
(HCG 97), an apparent merger that the optical data reveal to be a
superposition (RGH 80 and Abell 194), a linear cluster (Abell 194),
and three loose groups. These systems are a random and plausibly
representative subsample of the optically selected RASSCALS galaxy
groups.

\subsection{RGH80}

This group of galaxies was originally described by \nocite{Ramella89}{Ramella} {et~al.} (1989);
it was observed by ROSAT as well as ASCA, which yielded spectra with
emission-weighted temperatures $\approx 1.5$ keV and metallicities
$\approx 0.5$ solar \nocite{Hwang99}({Hwang} {et~al.} 1999). The XMM observations of this
group are described in detail by \nocite{Xue04}{Xue}, {B\"ohringer}, \& {Matsushita} (2004), who find that the data
are consistent with a monotonically decreasing metallicity
distribution.

Our new optical data suggests the presence of a second group in the
field of RGH 80. The X-ray peak to the SW of the central peak contains
another concentration of galaxies. These galaxies are also distinct
from the main peak. Figure~\ref{fig:rgh80hist} shows a histogram of
the rest-frame velocities of the galaxies within the central
12\arcmin. The 21 galaxies in this region exhibit a $490 \pm 10$ km
s\m\ void in velocity space.  The centers of the peaks themselves are
separated by $1080 \pm 103$ km s\m. If we split the two systems at the
midpoint of this void, we see that 7 of the 8 galaxies in the
higher-velocity peak reside within 30\arcsec\ (20 kpc) of the SW X-ray
peak. This lends support to the idea that the SW peak, separated from
the main peak by $156 \pm 5$ kpc, is a dynamically distinct group of
galaxies. The two groups are shown in Figure~\ref{fig:rgh80gemini}.

Several lines of reasoning suggests that the SW group has not
interacted with the main group. The void in the velocity histogram
($490$ km s\m) is comparable to the velocity dispersion of the group
overall ($\approx 600$ km s\m).  The size of this void argues against
a scenario in which the two groups have already interacted. If one
group were bound to and had already passed through the other, at least
a few galaxies within our detection limit ought to have been displaced
to occupy the velocity gap.

We can make this argument more rigorous by examining a two body model
first used by \nocite{Beers82}{Beers}, {Geller}, \& {Huchra} (1982) to study rich clusters of galaxies. In
this simplified scenario, the two groups are approximately point
masses in an orbit with zero angular momentum. Given a line-of-sight
velocity difference and a projected distance between the two bodies,
the model allows one to solve for $M(i)$, the total mass required to
bind the system as a function of the inclination angle $i$ between the
observer's line of sight and the orbital axis. We plot this function
in Figure~\ref{fig:mofi}. We also show the mass inferred from our
temperature measurements via the X-ray mass-temperature relation
\nocite{Finoguenov01b}({Finoguenov}, {Reiprich}, \& {B{\"  o}hringer} 2001). In the case of RGH 80, the main NE clump has an
emission-weighted temperature in the range 0.7--0.9 keV; for the SW
clump, the range is 0.5--0.7 keV, corresponding to an estimated mass
mass of 1.8--3.2 $\times 10^{13} M_\odot$ within $r_{500}$. (68\%
confidence, taking into account errors in the mass-temperature
relation as well as our own measurement errors).  To estimate the
remaining material within the virial radius $r_{200}$, we assume that
the matter distribution of both clumps can be described by a
\nocite{NFW}{Navarro} {et~al.} (1997) profile with transition radius $r_{500}$; we arrived at
this value by examining the mass profiles derived in the infall
regions of similar systems by \nocite{Rines03}{Rines} {et~al.} (2003), and converting their
best-fit NFW profiles to our $r_{500}$ reference frame. The virial
mass of the two clumps is then 1.3 times the mass at $r_{500}$, or
2.3--4.2 $\times 10^{13} M_\odot$. The minimum mass required to bind
the group is about 3 times larger than the X-ray mass inferred from
the mass-temperature relation.  In the context of the simple two-body
model, then, it is unlikely that the groups are bound.

The X-ray morphology leads to a similar conclusion. The maps of this
region show two distinct entropy minima, with no shocks indicating a
merging history.  Both the central clump and the southern clump are
embedded in separate low-entropy zones. It is more likely that the SW
pressure enhancement corresponds to a separate dark matter
concentration associated with the potential of the SW group, and less
likely that it corresponds to a shock. The spectral fits show that
there are two components consistent with being projected on top of
each other.  In the NE source, we observe a rise of the temperature
from 0.9 keV to 1.2 keV, and then in the outer regions back down again
to 0.7 keV. The metallicity is enhanced in the SW peak compared to the
rest of the image---it is distinct from its surroundings.
Thus the X-ray observations also support a scenario in
which the SW group is dynamically independent of the rest of the
system.

\resizebox{3in}{!}{\includegraphics{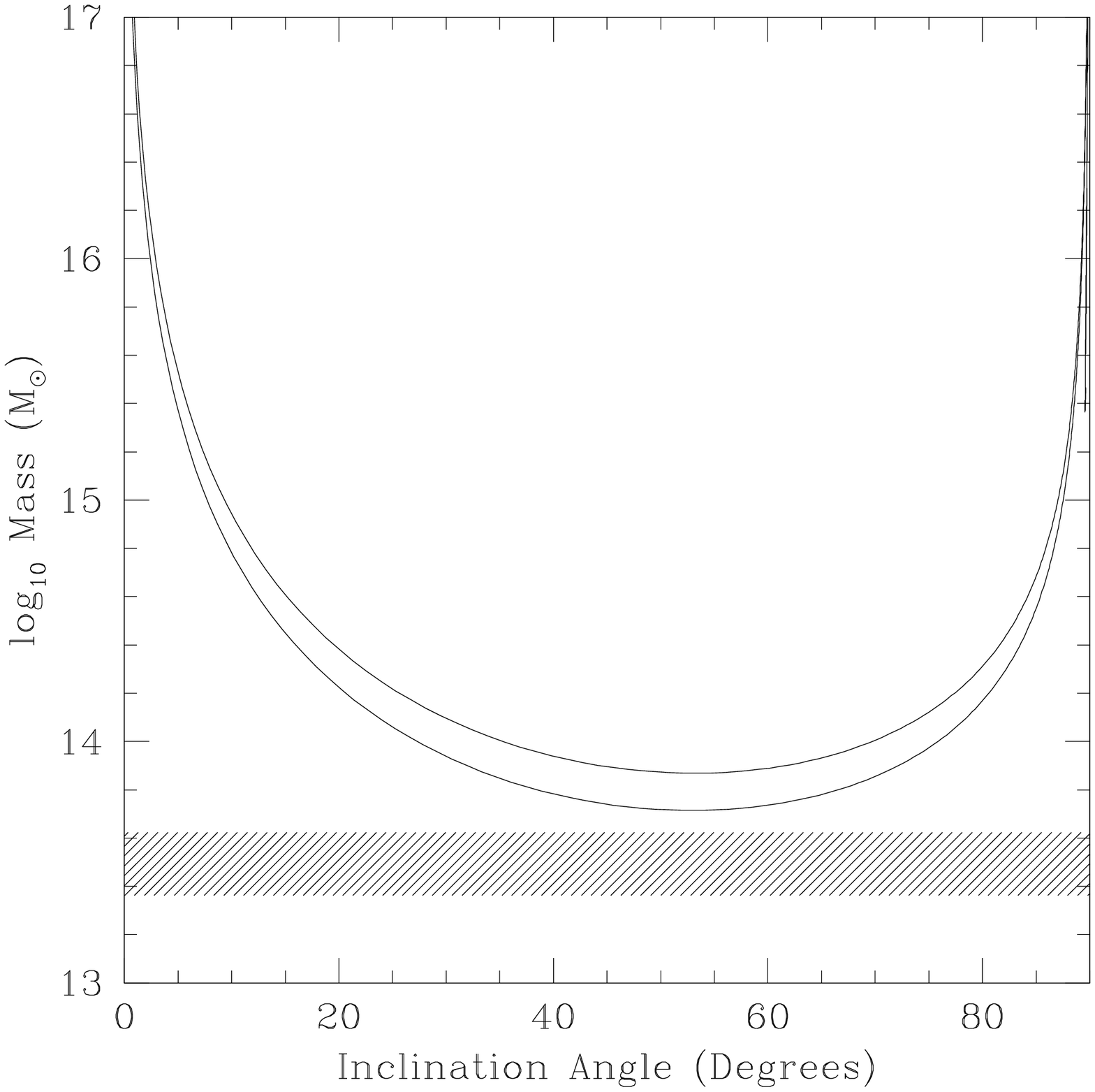}}
\figcaption{Simple two-body analysis of the NE and SW groups in RGH
80. The two solid lines indicate upper and lower limits on the minimum
total mass required to bind the two groups to each other. For
inclinations $i < 89$, only incoming bound solutions are possible; for
$i > 89^\circ$, both incoming and outgoing bound solutions are
possible, but the latter occupy only a miniscule region in the
solution space. The lower shaded region indicates the total mass of
the system derived from the X-ray mass-temperature relation. The X-ray
mass is too small to bind the groups by about a factor of 5.
\label{fig:mofi}}

\begin{figure*}
\begin{center}
\resizebox{6.5in}{!}{\includegraphics[angle=90]{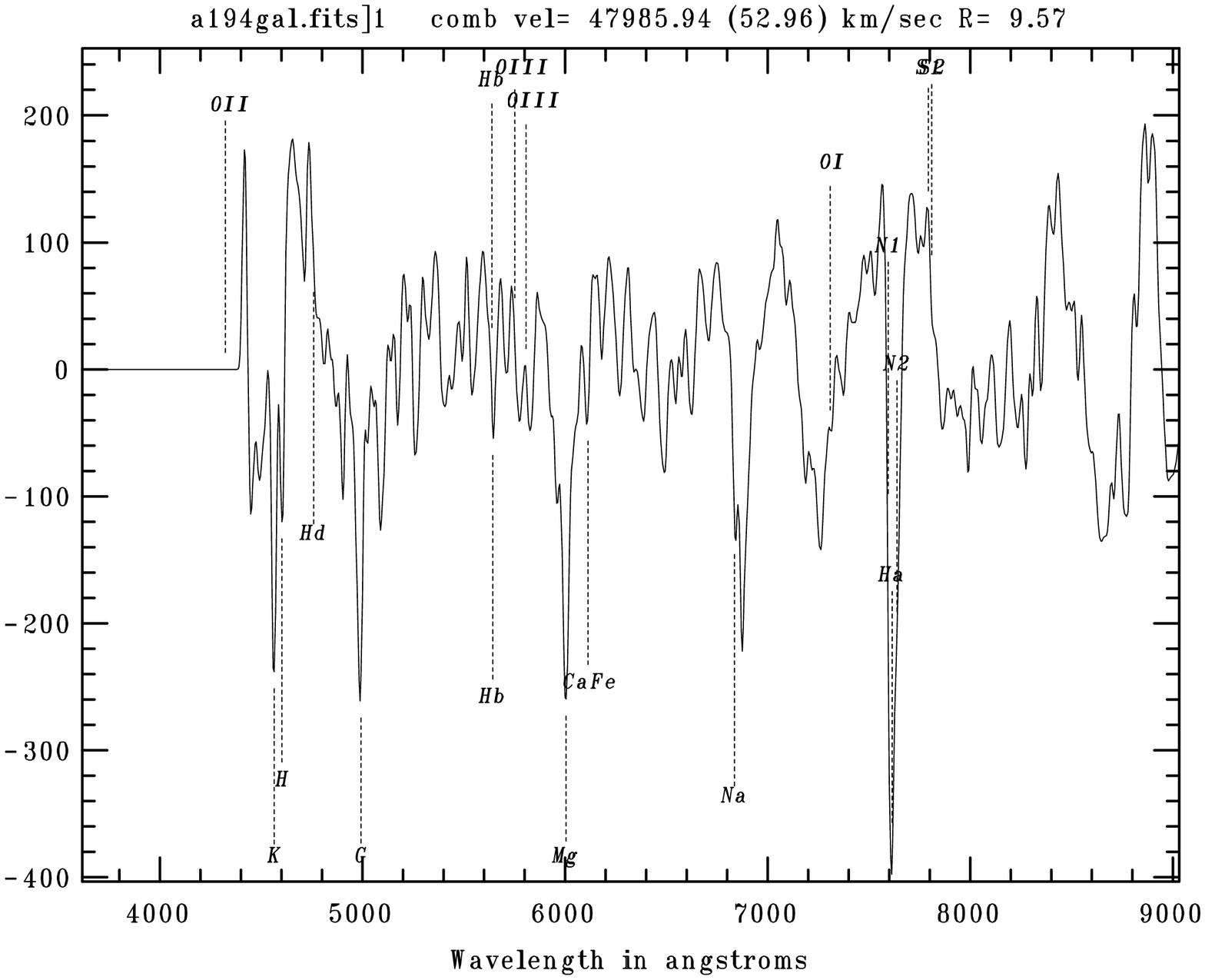}}
\figcaption{Continuum-subtracted spectrum of the galaxy associated
with the SW emission peak in the XMM-Newton observation of
Abell 194. The redshift of the galaxy indicates that it is a distant
background cluster, and not related to Abell
194. \label{fig:a194spec}}.
\end{center}
\end{figure*}
\begin{figure*}
\begin{center}
\resizebox{6.5in}{!}{\includegraphics{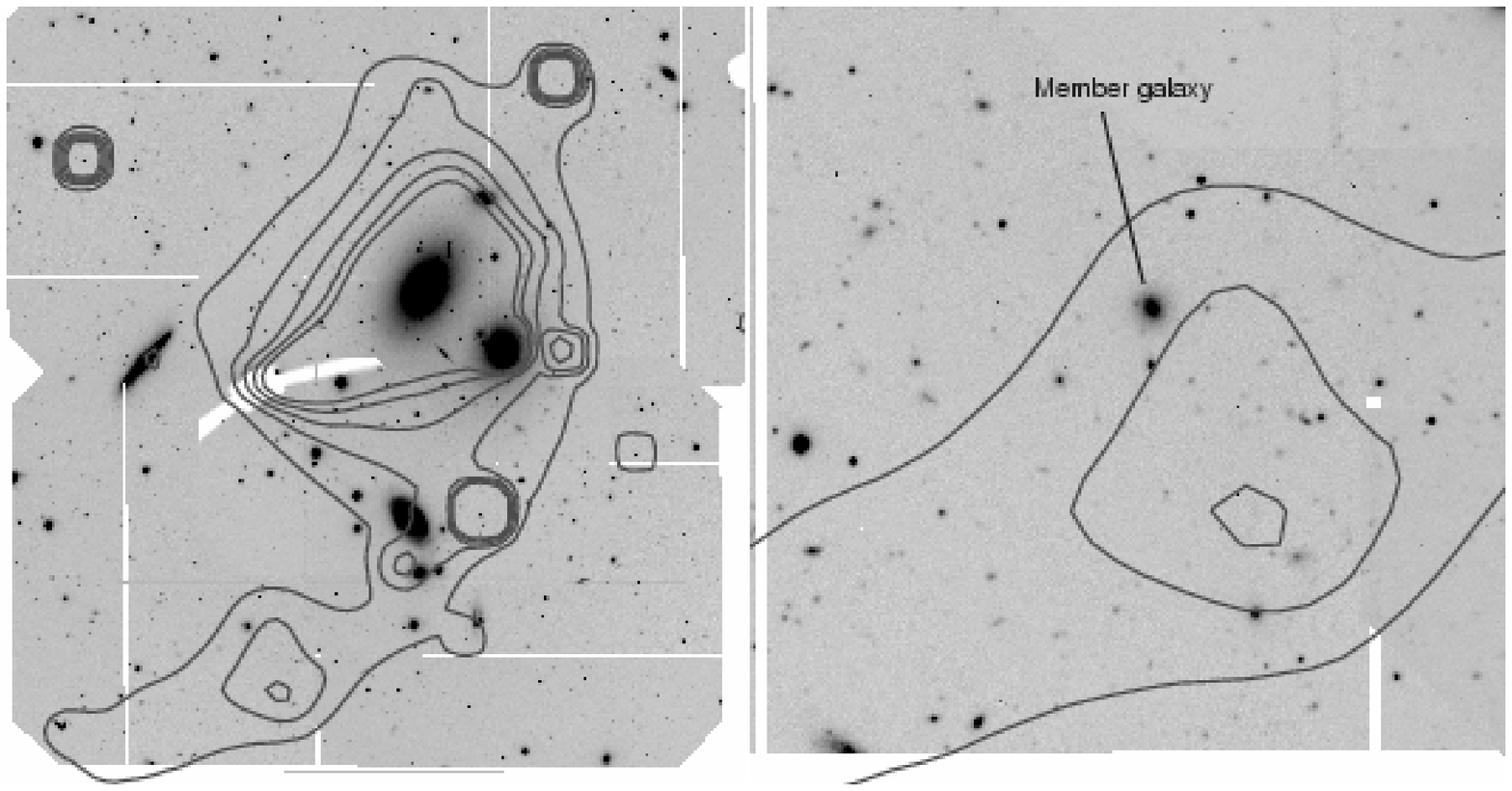}}
\figcaption{(\emph{left}) A 10 minute, $R$ band Gemini GMOS acquisition
image of HCG 97. X-ray surface brightness contours from our wavelet
decomposition technique are superimposed. The plume to the SE
(\emph{right}) is coincident with a single member galaxy. There are no
noticeable background clusters associated with the peak within the SE
extension.\label{fig:hcg97}}
\end{center}
\end{figure*}

\subsection{Abell 194}

Abell 194 was originally classified as a ``linear'' cluster of
galaxies by \nocite{Rood71}{Rood} \& {Sastry} (1971) because of its NE-SW alignment.  Our
XMM-Newton observation shows that the alignment also exists in
the X-ray. The first impression is that there are three peaks: one to
NE, one in the center, and one to the SW. The X-ray spectrum of the
central 15\arcsec of the NE peak is consistent with that of an active
galactic nucleus (AGN). 

It is however the southwesternmost X-ray peak that would seem to be of
particular interest, because \nocite{Knezek98}{Knezek} \& {Bregman} (1998) report that the galaxy at
its center has a redshift consistent with that of Abell 194. Yet this
galaxy is at least four magnitudes fainter than the galaxies occupying
the two other, bright X-ray emission peaks.  This would imply an
unreasonably large mass-to-light ratio of $\approx 1000$ for the SW
X-ray peak. Upon reanalyzing the \nocite{Knezek98}{Knezek} \& {Bregman} (1998) spectrum (Figure
\ref{fig:a194spec}), we find that its recession velocity is $47900 \pm
140$ km s\m. Also, the X-ray spectrum corresponding to the SW region
is better fit by a thermal plasma at redshift 0.15 than it is by a
plasma at redshift 0.017. The galaxy atop the SW X-ray peak is
therefore the brightest member of a more distant cluster, and the peak
is not associated with Abell 194, but corresponds to the X-ray
atmosphere of the cluster at $z = 0.15$. This result, together with
the result for RGH 80, stresses the importance of projection effects
when considering the properties of clusters and groups of galaxies
\nocite{Cen97}({Cen} 1997).

Thus, there are only two distinct halos in Abell 194, both exhibiting
an outward rise in temperature from 0.7 keV in the center to 1.6 keV
in the outskirts. The velocity distributions of the galaxies within
each X-ray peak are indistinguishable; as a result it is likely
that the two clumps are bound.

The entropy profile of Abell 194 seems to resist an azimuthally
averaged description (\S~\ref{sec:profiles}). The cluster registers a
larger $\chi^2$ than the other seven systems in our sample, regardless
of the entropy model. Given that it is by far the least axisymmetric
cluster in our sample, the disagreement is hardly a surprise. However,
it is surprising that the pressure profile is fully consistent with
the self-similar model fit to the rest of the sample.

\subsection{HCG 97}

HCG 97 is a group of galaxies originally discussed in the
\nocite{Hickson82}{Hickson} (1982) catalog of compact systems. \nocite{Ebeling94}{Ebeling} {et~al.} (1994) first
reported X-ray emission in the ROSAT All-Sky Survey for this
group.  

Our analysis of the XMM-Newton observation reveals a complex structure
in this interacting system, with a large gas tail stretching toward
the SE. This tail is a source of low-entropy gas and may account for
the deficit seen in the entropy profile at around $0.5 r_{500}$. The
Gemini observation reveals a single member galaxy inhabiting the SE
extension plume; a close examination of the GMOS acquisition image
(Figure~\ref{fig:hcg97}) reveals no background clusters that could be
responsible for the X-ray emission in that region.

If the SE plume is associated with the cluster, it may be be a relic
of the infall of one of the brighter interacting galaxies at the group
center---or it may be yet another chance superposition from a cluster
not visible in the Gemini image. The temperature of the plume is
consistent with that of its surroundings; i.e., the plume is not
apparent in the temperature map (Figure~\ref{fig:snkt}). However, the
plume is clearly present in the entropy, pressure, and density
distributions.  Neither the X-ray and the optical data are at this
point of sufficient quality to determine whether the plume has been
stripped from one of the central galaxies, or whether it is a
background structure unrelated to the main group.

Substructure notwithstanding, the azimuthally averaged pressure and
entropy distributions in HCG 97 are consistent with the self-similar
power laws fitted to the sample as a whole.

\subsection{SS2b153}

SS2b153 is a remarkably round system, a stark contrast to the
complexity of the other groups. Because the only redshifts available
in its region are from the nearly decade-old Las Campa\~nas Redshift
Survey \nocite{Shectman96}({Shectman} {et~al.} 1996), very few galaxies are known to be members.

The azimuthally averaged pressure and entropy distributions in this
system are consistent with the self-similar broken power law fit to
the sample as a whole, except for the innermost data point. The
pressure at $\approx 10$ kpc disagrees with the self-similar law by a
factor of 2. SS2b153 is the most nearby group in the sample, and it is
possible that the pressure enhancement represents the contribution of
the gas bound to the central galaxy, NGC 3411, to the total
pressure. Similar enhancements---associated with the central galaxies
of groups within $z \approx 0.01$---have been observed in other
systems \nocite{Helsdon01}({Helsdon} {et~al.} 2001).

\subsection{NGC 5129 Group}

The ``plume'' to the NE seems to be indistinguishable from its
environment to within the errors. The entire central region seems
roughly isothermal, with a temperature 0.7 keV. The entropy profile of
the NGC 5129 Group does not reveal any deviations from the
self-similar models, and its pressure profile also appears consistent.

\subsection{Abell 2634}

The XMM-Newton observation of Abell 2634 was plagued by long
periods of high X-ray background, which did not allow us to extract as
much spectroscopic information as would have otherwise been possible
for this bright cluster.  The radio source 3C 465 coincides with the
central galaxy \nocite{Owen76,Sakelliou99}({Owen} \& {Rudnick} 1976; {Sakelliou} \& {Merrifield} 1999); it exhibits a powerful jet
that interacts the X-ray emitting gas as far out as 100 kpc from the
cluster center. For this reason, our entropy and pressure profiles
miss the central section of the cluster entirely.

\subsection{NRGb184}

The group has a north/south elongation in surface brightness.  This
elongation is less pronounced in the pressure map.  There is a $\sim
20\%$ pressure difference between the regions directly to the SE and
NW of the group center. The temperature rises fairly quickly from 0.6
keV in the very central region to about 1.5 keV.  The N-S extension
that is visible in the surface brightness map is not present in
temperature map. On both the entropy and pressure plots, the
properties of the gas in NRGb184 appear typical.

\begin{tabular}{c}
\resizebox{3.5in}{!}{\includegraphics[angle=90]{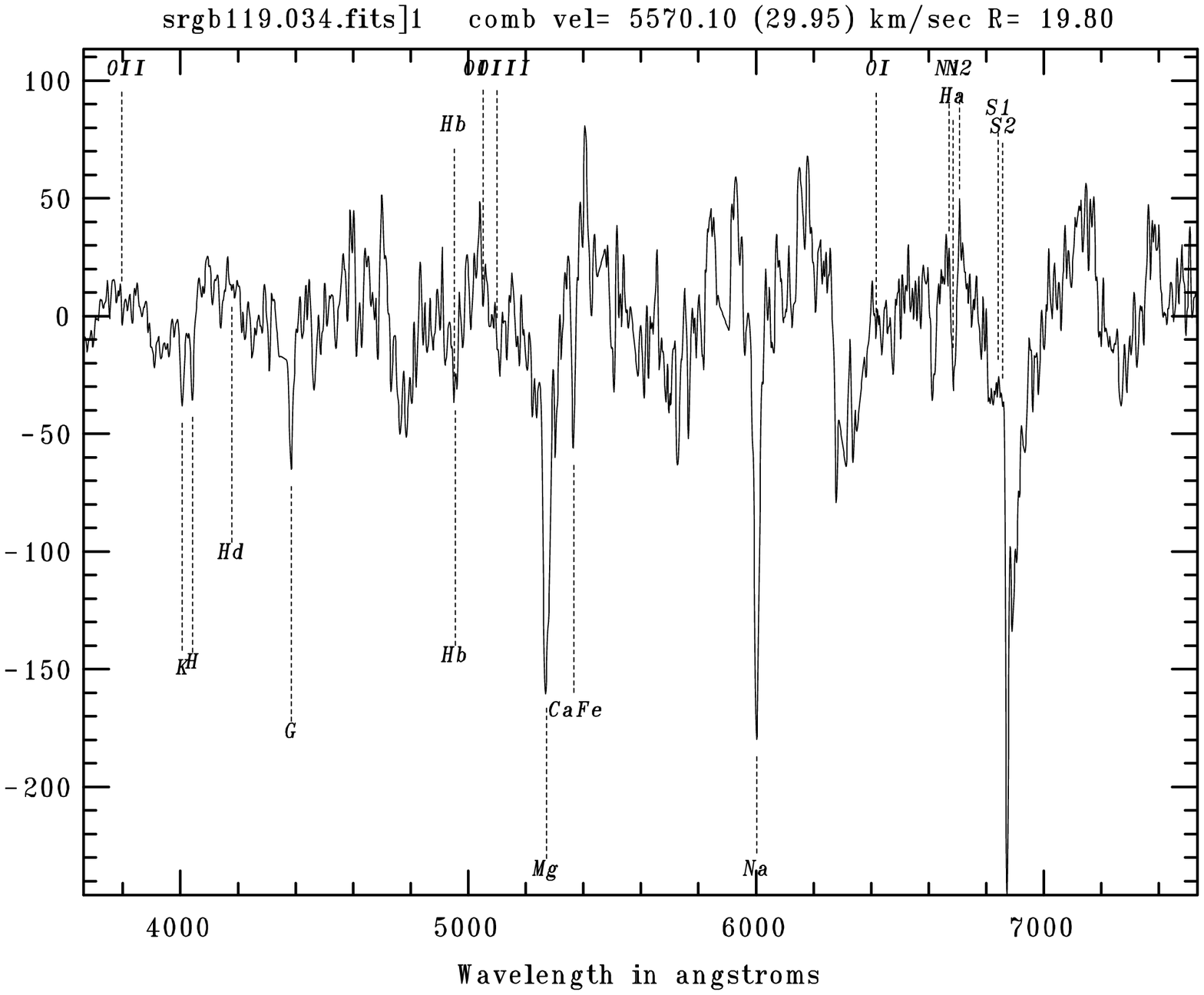}}\\
\resizebox{3.5in}{!}{\includegraphics[angle=90]{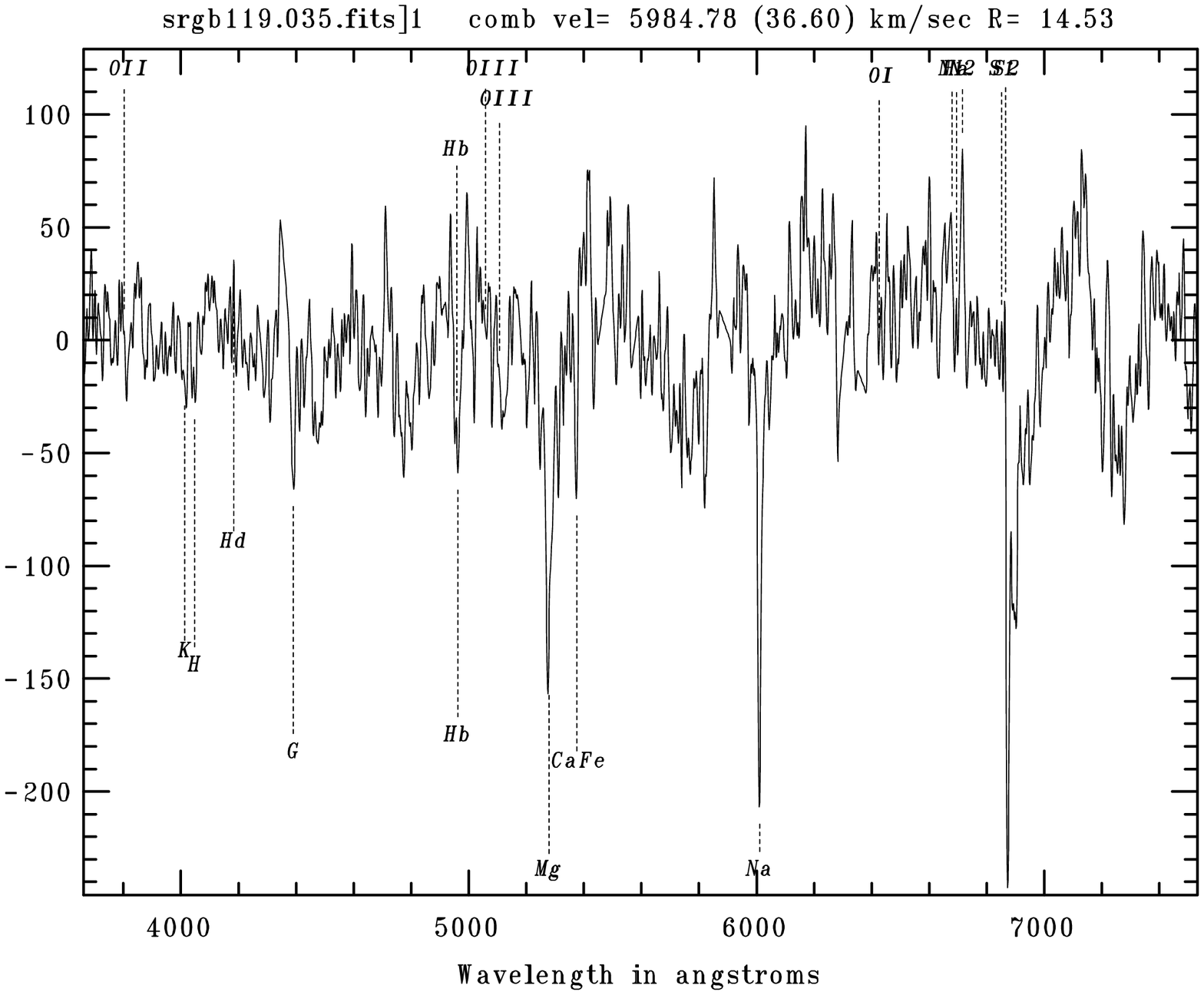}}
\end{tabular}
\figcaption{Continuum-subtracted spectra of the western (\emph{top})
and eastern (\emph{bottom}) pair of central galaxies in SRGb119.
\label{fig:srgb119}.}

\subsection{SRGb119}

In general, the temperature increases from 0.7 keV towards the center
to 1.4 keV towards the exterior. There are too few counts to 
constrain the iron abundance robustly.
--
There is an interesting unresolved source to the east of the very
center of the group. It is a member galaxy that is $\sim 2$ mag
fainter than the brightest group galaxy. The two galaxies have a
large velocity difference (450 km s\m).
Despite being fainter, the galaxy to the E has a higher temperature
than the central source, $1.6 \pm 0.2$ keV compared to $0.7 \pm 0.2$
keV.  The optical spectra of both the central galaxies show no
evidence of AGN emission (Figure~\ref{fig:srgb119}). The higher X-ray
temperature for the source to the east is therefore a puzzle. Our
analysis rules out a significant contribution coming from low-mass
X-ray binaries (LMXBs): the spectrum is inconsistent with the $\gtrsim
5$ keV temperatures expected from these objects \nocite{Sarazin01}({Sarazin}, {Irwin}, \& {Bregman} 2001). At
the same time, however, a power law spectrum with index $2.9 \pm 0.4$
provides a better fit than a thermal plasma. The power law spectrum
has a total luminosity of $9\times10^{40}$ ergs/s in the 0.5--2 keV
band. The source of this spectrum could be a veiled AGN such as the
one observed in NGC 1291 by \nocite{Irwin02}{Irwin}, {Sarazin}, \& {Bregman} (2002).

The pressure and entropy distribution in SRGb119 presents a puzzle
similar to that of Abell 194. By itself, the azimuthally averaged
entropy profile of SRGb119 is not a good fit to the simple empirical
models in \S\ref{sec:profiles}. Of the eight groups in our sample,
SRGb119 makes the second largest contribution to the reduced $\chi^2$
(after Abell 194), regardless of the shape of the model. However,
similarly to Abell 194, SRGb119 is quite consistent with the
self-similar pressure models. Thus we find that the two groups with
the more discrepant entropy distributions are nevertheless similar to
the rest of the sample when we consider the pressure profiles.

\section{Conclusion}
\label{sec:conclusion}

We use new and previously unpublished archival XMM-Newton data
to examine the properties of the X-ray emitting medium in eight groups
of galaxies from the RASSCALS survey. In addition to the X-ray data,
we collect redshifts for the group members from the literature. We
also report redshifts from a Gemini North Observatory survey of the
groups RGH 80, HCG 97, and NRGb184.

The sample consists of groups with X-ray luminosities $\lesssim
10^{43}$ erg s\m, and was chosen based on availability in the
XMM-Newton data archive. The groups in the sample exhibit a wide
variety of X-ray and optical morphologies, from seemingly perfectly
spherical systems (SS2b153) to linear clusters (Abell 194). Many of
the systems exhibit certain or possible superposition, in both the
X-ray and optical, with unrelated background galaxies or clusters of
galaxies. We use the Gemini North optical data to argue that RGH 80 is
a superposition of two dynamically unrelated groups.

Despite the morphological incongruity of the groups, they exhibit
regularity in their azimuthally averaged pressure and entropy
profiles. Self-similar broken power laws can describe the spatial
distribution of both these quantities. For the entropy, which
increases with distance from the center, we find an inner slope
$\alpha_s = 0.92^{+0.04}_{-0.05}$ and outer slope $\gamma_s =
0.42^{+0.05}_{-0.04}$ (68\% confidence). These results argue against
the existence of an isentropic ``floor'' at the center of X-ray
emitting groups, and against theoretical models that describe the
entropy distribution as a single power law for $r \gtrsim 0.3 r_{500}$.

The decreasing pressure profiles are also well-fitted by a
self-similar broken power law. The inner and the outer slopes are
$\alpha_p = -0.78^{+0.04}_{-0.03}$ and $\gamma_p = -1.7^{+0.1}_{-0.3}$
for the entire sample. The regularity and similarity of the pressure
profiles suggests that the groups reside in dark matter halos that are
similar in structure.

We have made use of the NASA/IPAC Extragalactic Database (NED) which
is operated by the Jet Propulsion Laboratory, California Institute of
Technology, under contract with the National Aeronautics and Space
Administration. We thank Patricia Knezeck for providing us with the
spectrum of the galaxy in the SW X-ray peak in Abell 194, and the
anonymous referee for comments that improved the
paper. J. P. H. thanks the Alexander von Humboldt Foundation for
support in Garching.  M. J. G.  was supported by the Smithsonian
Institution. A. F. acknowledges support from BMBF/DLR under grant 50
OR 0207, from MPG, and from NASA grant NNG04GH226.  A. M. was
supported by NASA through a Chandra Postdoctoral Fellowship Award
issued by the Chandra X-ray Observatory Center, which is operated by
the Smithsonian Astrophysical Observatory for and on behalf of NASA
under contract NAS 8-39073.

\bibliography{}

\newcommand{\tna}{\tablenotemark{a}}
\begin{deluxetable}{lrrrrrccc}
\tablewidth{0in}
\tabletypesize{\footnotesize}
\tableheadfrac{0.35}
\tablecaption{Group Sample \label{tbl:sample}}
\tablehead{\colhead{Primary ID} & \colhead{$\alpha_{2000}$} & 
\colhead{$\delta_{2000}$} & \colhead{$N$} &\colhead{$c \bar{z}$} & 
\colhead{$\sigma_\mr{los}$} & \colhead{Other ID} 
& \colhead{Central Galaxy} & \colhead{Sources}}
\startdata
Abell 194 & 01:25:48.1 & -01:23:06 &106 & $5316^{+53}_{-53}$ & $550^{+89}_{-86}$ & SRGb103 & \tna     & R(106) \\
SRGb119    & 01:56:20.8 & +05:37:40 & 61 & $5525^{+52}_{-48}$ & $416^{+34}_{-33}$  & WBL 61 & NGC  741 & M(21), Z(30) \\
SS2b153    & 10:50:26.3 & -12:50:43 &  5 & $4747^{+59}_{-69}$ & $161^{+19}_{-71}$  & USGC S152 & NGC 3411 & L(14)  \\
NRGb184   & 12:08:05.5 & +25:14:14 & 33 & $6700^{+46}_{-62}$ & $390^{+40}_{-37}$ &\nodata & UGC 07115& G(8), M(52) \\
RGH80   & 13:20:14.8 & +33:08:35 & 58 &$11047^{+74}_{-80}$ & $602^{+63}_{-61}$ &  NRGs241   & NGC 5098 & G(14), M(44) \\
NGC 5129 & 13:24:09.6 & +13:58:51 & 33 & $6952^{+46}_{-44}$ & $283^{+29}_{-31}$ & NRGb244 & NGC 5129 & M(19), Z(15) \\
Abell 2634& 23:38:29.2 & +27:01:54 & 69 &  $9391^{+73}_{-72}$ & $568^{+39}_{-43}$ & 3C 465 / SRGs040 & NGC 7720 & P(69) \\
HCG 97    & 23:47:23.1 & -02:18:07 & 37 & $6758^{+62}_{-49}$ & $383^{+50}_{-52}$ & SS2b312 & IC 5357  & G(11), M(26) \\ 
\enddata 
\tablecomments{$N$ is the number of known members within 2\ Mpc of
the X-ray center of the group; $c z$ is the mean redshift of the group
multiplied by the speed of light in km s\m; $\sigma_{\rm los}$ is the velocity
dispersion of the group in km s\m. Also shown are other names of each
group, as well as the common name of the central galaxy. The redshifts
are taken from the following sources: \nocite{Mahdavi04}{Mahdavi} \& {Geller} (2004, M);
\nocite{Shectman96}{Shectman} {et~al.} (1996, L); \nocite{Zabludoff98}{Zabludoff} \& {Mulchaey} (1998, Z); \nocite{Ramella02}{Ramella} {et~al.} (2002, U);
\nocite{Rines03}{Rines} {et~al.} (2003, R); \nocite{Pinkney93}{Pinkney} {et~al.} (1993, P); and Gemini (G) redshifts 
measured in this
work. The number after each letter indicates the number of unique
galaxies that were measured in that work.}
\tablenotetext{a}{The ``chain cluster'' Abell 194 does not have a dominant galaxy.}
\end{deluxetable}

\begin{deluxetable}{rrrrr}
\tablewidth{0in}
\tabletypesize{\footnotesize}
\tableheadfrac{0.35}
\tablecaption{New Optical Data \label{tbl:gemini}}
\tablehead{\colhead{$\alpha_{2000}$} & 
\colhead{$\delta_{2000}$} & \colhead{$c z$} &\colhead{$\epsilon_{c z}$} & \colhead{Type\tna}}
\startdata
12:07:44.31 & +25:14:48.5 &    41241 &  8 & E \\
12:07:45.67 & +25:14:36.9 &    26338 &  5 & E \\
12:07:46.99 & +25:15:51.5 &     6468 & 16 & A \\
12:07:47.71 & +25:11:13.2 &    62187 & 102 & A \\
12:07:49.53 & +25:12:35.9 &    29447 &  9 & E \\
12:07:54.08 & +25:16:58.9 &   155457 & 22 & E \\
12:07:55.33 & +25:14:00.2 &    89416 & 30 & A \\
12:07:57.64 & +25:16:43.3 &   126564 & 12 & E \\
12:07:58.16 & +25:10:17.8 &   112008 & 38 & A \\
12:07:59.33 & +25:11:45.9 &     5896 & 59 & A \\
12:08:00.12 & +25:15:14.9 &    85701 & 25 & E \\
12:08:01.94 & +25:15:43.3 &   100206 & 40 & A \\
12:08:02.23 & +25:11:18.2 &     7096 & 36 & A \\
12:08:04.01 & +25:16:30.1 &   134066 & 36 & A \\
12:08:04.68 & +25:19:10.0 &   142108 & 24 & E \\
12:08:06.12 & +25:16:02.9 &     5813 & 23 & A \\
12:08:06.23 & +25:13:31.6 &   160608 & 44 & A \\
12:08:07.16 & +25:15:36.9 &     6502 & 19 & A \\
12:08:07.44 & +25:12:51.5 &     6964 & 38 & A \\
12:08:08.16 & +25:14:44.3 &    89252 & 22 & A \\
12:08:08.86 & +25:13:05.2 &    25449 & 19 & A \\
12:08:10.42 & +25:16:03.3 &   197516 & 30 & E \\
12:08:11.97 & +25:11:21.9 &    24130 & 12 & E \\
12:08:12.37 & +25:18:04.3 &   189232 & 42 & E \\
12:08:13.18 & +25:15:40.3 &    92668 & 46 & A \\
12:08:14.87 & +25:09:38.8 &   153102 & 27 & E \\
12:08:15.42 & +25:13:23.3 &   155169 & 31 & E \\
12:08:15.86 & +25:16:18.6 &    79235 & 14 & E \\
12:08:17.88 & +25:14:32.2 &    78955 & 16 & E \\
12:08:18.85 & +25:13:46.3 &    78683 &  9 & E \\
12:08:19.19 & +25:15:47.2 &   189886 & 34 & E \\
12:08:19.37 & +25:13:57.9 &     6912 & 60 & A \\
12:08:21.44 & +25:09:40.9 &   110299 & 14 & E \\
12:08:22.19 & +25:17:58.6 &    25738 & 13 & E \\
12:08:22.26 & +25:18:57.8 &     6827 & 26 & A \\
12:08:22.88 & +25:09:48.9 &   110235 & 17 & E \\
12:08:25.33 & +25:16:20.9 &    78222 & 15 & E \\
13:20:00.91 & +33:09:11.9 &   129328 & 61 & A \\
13:20:07.44 & +33:07:00.8 &    10991 & 60 & A \\
13:20:08.12 & +33:10:38.0 &    10999 & 29 & A \\
13:20:08.32 & +33:06:13.0 &    12317 & 55 & A \\
13:20:14.00 & +33:07:34.0 &    10483 & 51 & A \\
13:20:14.70 & +33:12:17.6 &    10594 & 27 & A \\
13:20:15.01 & +33:05:38.1 &    11922 &  6 & E \\
13:20:15.05 & +33:11:57.3 &    67931 & 28 & E \\
13:20:15.99 & +33:04:38.3 &    29945 & 11 & E \\
13:20:17.45 & +33:08:54.4 &    80238 & 81 & A \\
13:20:17.76 & +33:12:12.1 &    35837 & 10 & E \\
13:20:18.79 & +33:10:17.9 &   105272 & 67 & A \\
13:20:18.80 & +33:09:55.0 &   104821 & 12 & E \\
13:20:19.03 & +33:08:03.9 &    10725 & 33 & A \\
13:20:20.01 & +33:13:30.1 &    11005 & 14 & E \\
13:20:21.27 & +33:04:41.0 &    10283 &  6 & E \\
13:20:23.04 & +33:09:17.5 &    11001 & 65 & A \\
13:20:23.59 & +33:06:40.3 &    98242 & 34 & A \\
13:20:25.40 & +33:11:24.2 &     5460 & 20 & E \\
13:20:27.84 & +33:08:27.4 &    10735 & 27 & A \\
13:20:31.15 & +33:06:08.8 &    68601 & 29 & A \\
13:20:31.69 & +33:07:14.3 &    11008 & 62 & A \\
13:20:33.30 & +33:05:59.0 &    11714 &  5 & E \\
13:20:33.40 & +33:12:00.0 &   141341 & 29 & E \\
13:20:35.19 & +33:05:40.7 &   133709 & 41 & E \\
13:20:36.30 & +33:09:45.0 &    20144 & 17 & E \\
23:47:05.53 & -02:14:30.6 &    42715 &  4 & E \\
23:47:06.20 & -02:14:10.4 &    35609 & 21 & E \\
23:47:09.34 & -02:15:37.6 &     7568 & 48 & A \\
23:47:09.37 & -02:22:40.8 &    89460 & 22 & E \\
23:47:09.92 & -02:18:54.8 &   140450 & 18 & E \\
23:47:10.10 & -02:15:11.8 &     7591 &  6 & E \\
23:47:10.65 & -02:19:38.7 &   116770 & 23 & E \\
23:47:11.65 & -02:22:34.3 &   152617 & 46 & A \\
23:47:14.70 & -02:21:53.4 &    45685 & 14 & E \\
23:47:15.40 & -02:21:35.5 &   115631 & 54 & A \\
23:47:15.60 & -02:20:50.4 &   115784 & 25 & A \\
23:47:16.20 & -02:19:09.0 &    96754 &  7 & E \\
23:47:18.90 & -02:24:12.8 &    77719 & 10 & E \\
23:47:19.40 & -02:21:19.0 &    89411 & 26 & E \\
23:47:19.90 & -02:16:50.4 &     6665 & 24 & A \\
23:47:19.94 & -02:16:34.0 &    63146 & 14 & E \\
23:47:20.10 & -02:18:37.6 &    88188 & 25 & A \\
23:47:20.30 & -02:22:26.8 &     7247 &  4 & E \\
23:47:20.94 & -02:17:32.5 &   124893 & 24 & E \\
23:47:21.00 & -02:19:28.5 &   140106 & 32 & E \\
23:47:21.70 & -02:17:32.2 &    29293 & 11 & E \\
23:47:22.00 & -02:18:53.4 &    29389 & 53 & A \\
23:47:22.00 & -02:21:12.6 &     6033 & 48 & A \\
23:47:22.62 & -02:18:14.6 &    68783 & 39 & A \\
23:47:22.80 & -02:15:25.0 &    88275 & 30 & A \\
23:47:23.30 & -02:21:47.0 &     5991 &  7 & E \\
23:47:23.88 & -02:16:02.6 &   111777 & 28 & E \\
23:47:24.72 & -02:16:35.6 &   152587 & 14 & E \\
23:47:24.96 & -02:16:24.1 &   124710 & 38 & A \\
23:47:25.00 & -02:19:35.5 &     5837 & 37 & A \\
23:47:26.32 & -02:16:44.4 &    55222 &  8 & E \\
23:47:26.60 & -02:21:11.7 &     6778 & 36 & A \\
23:47:26.60 & -02:21:11.8 &     6519 & 25 & E \\
23:47:26.60 & -02:21:11.8 &     6698 &  6 & E \\
23:47:27.36 & -02:17:22.9 &   169001 & 32 & E \\
23:47:28.36 & -02:24:00.3 &   169622 & 27 & A \\
23:47:28.81 & -02:20:12.7 &     -117 & 45 & A \\
23:47:29.50 & -02:18:04.8 &    68533 & 16 & E \\
23:47:29.50 & -02:18:05.0 &    68523 & 12 & E \\
23:47:29.60 & -02:22:56.3 &   169789 &  5 & E \\
23:47:30.00 & -02:23:30.8 &    41439 & 13 & E \\
23:47:30.23 & -02:16:51.4 &   125224 & 26 & A \\
23:47:30.50 & -02:22:01.6 &   168233 & 30 & E \\
23:47:30.80 & -02:19:08.3 &    45611 & 30 & A \\
23:47:31.20 & -02:20:28.2 &    42185 &  8 & E \\
23:47:31.20 & -02:20:28.2 &    42206 & 45 & A \\
23:47:32.10 & -02:18:18.3 &    72077 & 16 & E \\
23:47:32.40 & -02:22:29.3 &     6879 & 66 & A \\
23:47:32.47 & -02:16:24.8 &   115977 & 39 & A \\
23:47:34.65 & -02:18:48.5 &   185245 & 28 & E \\
23:47:35.20 & -02:21:39.1 &    84656 & 17 & E \\
23:47:35.74 & -02:15:30.1 &   135273 & 37 & E \\
23:47:36.00 & -02:24:10.7 &   129738 & 26 & A \\
23:47:36.83 & -02:17:13.0 &    69053 &  9 & E \\
23:47:38.20 & -02:20:38.0 &   183269 & 24 & E \\
23:47:38.47 & -02:16:28.4 &   106316 & 28 & E \\
23:47:39.27 & -02:14:38.2 &   169012 & 10 & E \\
23:47:40.04 & -02:18:57.9 &   139059 & 11 & E \\
23:47:40.90 & -02:21:01.5 &    68338 & 12 & E \\
23:47:42.40 & -02:23:47.5 &    66326 & 19 & E \\
23:47:42.64 & -02:14:33.1 &   115187 & 21 & A \\
23:47:42.70 & -02:21:51.2 &   129539 & 42 & A \\
23:47:43.80 & -02:17:49.6 &   138877 & 21 & E \\
\enddata
\tablecomments{$c z$ denotes the redshift of the object
multiplied by the speed of light; $\epsilon_{c z}$ is
the $1 \sigma$ uncertainty in $ c z$; ``type''
indicates whether the dominant features in the spectrum
are absorption (``A'') or emission (``E'') lines.}
\end{deluxetable}

\newcommand{\mcf}[1]{\multicolumn{4}{c}{#1}}

\begin{deluxetable}{lccccr}
\tablewidth{0in}
\tablecaption{XMM Observational Information \label{tbl:xmmprops}}
\tablehead{ Name &  Obs. ID  & net exp & filter  & Orbit   &frame time\\
      &             &             ksec   &       &       &ms}
\startdata
Abell 194   &0136340101 &12.9& Thin  & 557& 73\\
SRGb119     &0153030701 & 3.9& Thin  & 745& 73\\
SS2b153     &0146510301 &18.5& Thin  & 555& 73\\
NRGb184     &0151400201 & 8.6& Thin  & 634& 73\\
RGH 80      &0105860101 &26.2& Thin  & 563&199\\
NGC 5129    &0108860701 &12.4& Medium& 472& 73\\
Abell 2634  &0002960101 & 4.0& Medium& 464& 73\\
HCG 97      &0152860101 &20.9& Thin  & 559& 73\\
\enddata
\end{deluxetable}

\begin{deluxetable}{lcccccc}
\tablewidth{0in}
\tablecaption{IGM Properties of the Groups between $0.1-0.5 r_{500}$. \label{t:adv}}
\tablehead{\colhead{Group} & \colhead{$r_{500}$} & \colhead{$\langle k T\rangle$}  & 
\colhead{$\langle Z\rangle$}  & \colhead{$\langle S \rangle$}  & \colhead{$\langle P\rangle$}\\
\colhead{} &
\colhead{kpc} &
\colhead{keV} &
\colhead{$Z_\odot$} &
\colhead{keV cm$^2$} &
\colhead{$10^{-12}$ ergs cm$^{-3}$} }
\startdata
Abell 194  &521& $1.36\pm0.04$&$0.19\pm 0.03$&$260.\pm15.1$&$0.89\pm0.06$ \\ 
SRGb119&520& $1.36\pm0.07$&$0.55\pm 0.27$&$402.\pm98.7$&$0.46\pm0.13$ \\ 
SS2b153&407& $0.65\pm0.02$&$0.15\pm 0.01$&$109.\pm 6.8$&$0.47\pm0.03$ \\ 
NRGb184&488& $1.26\pm0.07$&$0.07\pm 0.02$&$192.\pm13.0$&$1.11\pm0.09$ \\ 
RGH 80  &455& $1.04\pm0.02$&$0.16\pm 0.02$&$107.\pm 6.1$&$1.67\pm0.10$ \\ 
NGC 5129&434& $0.95\pm0.05$&$0.39\pm 0.12$&$178.\pm26.2$&$0.60\pm0.13$ \\ 
Abell 2634 &639& $2.07\pm0.11$&$0.20\pm 0.05$&$259.\pm16.3$&$2.42\pm0.16$ \\ 
HCG 97  &420& $0.89\pm0.02$&$0.17\pm 0.03$&$129.\pm 8.6$&$0.79\pm0.06$ 
\label{tbl:xprops} 
\enddata
\end{deluxetable}

\begin{deluxetable}{rcrc}
\tablewidth{0in}
\tablecaption{Self-similar Pressure and Entropy Profile Fits \label{tbl:results}}
\tablehead{\mc{Entropy Fits} & \mc{Pressure Fit} }
\startdata
\mcf{Entire Sample} \\ 
Goodness-of-fit & 0.0095 ($\chi^2/\nu = 93/62$) & & 0.15 ($\chi^2/\nu = 73/62$)\\
    $\alpha_s$ & 0.87--0.96 &-$\alpha_p$ & 0.75--0.82\\
    $\gamma_s$  & 0.38--0.47 &-$\gamma_p$  & 1.4--1.8  \\
\mcf{Abell 194} \\ 
$r_s$    & $> 0.17 $      &$r_p$   & 0.16--0.20 \\
$S_0$    & 120--140       &$P_0$   & 17--20     \\
\mcf{SRGb119} \\			  		  
$r_s$    & $> 0.14$      &$r_p$    & 0.042--0.056\\
$S_0$    & 150--190      &$P_0$    & 17--26\\
\mcf{SS2b153} \\ 			  		  
$r_s$    & 0.054--0.069  &$r_p$    &0.014--0.047\\
$S_0$    & 100--120      &$P_0$    & 19--23 \\
\mcf{NRGb184} \\			  		  
$r_s$    & 0.062--0.083  &$r_p$    &$> 0.19$\\
$S_0$    & 170--200      &$P_0$    & 12--15 \\
\mcf{NGC 5129} \\			  		  
$r_s$    & 0.062--0.10   &$r_p$    &0.042--0.070\\
$S_0$    & 160--200      &$P_0$    & 11--14 \\
\mcf{RGH 80} \\ 			  		  
$r_s$    & 0.048--0.068  &$r_p$    & 0.042--0.096\\
$S_0$    & 110--130      &$P_0$    & 26--31 \\
\mcf{Abell 2634} \\
$r_s$    & $< 0.062$     &$r_p$    & 0.12--0.14 \\
$S_0$    & 240--600      &$P_0$    & 75--96     \\
\mcf{HCG 97} \\ 			  		  
$r_s$    & 0.021--0.030  &$r_p$    &0.083--0.098\\
$S_0$    & 180--250      &$P_0$    & 14--16\\
\enddata
\tablecomments{
the 68\% confidence intervals take into account correlations with all
the other parameters. The transition radii $r_s$ and $r_p$ are in
units of Mpc; $S_0$ is in units of keV cm$^{2}$, and $P_0$ is in units
of $10^{-12}$ erg cm$^{-3}$. }
\end{deluxetable}
 
\end{document}